\documentclass[12pt]{article}
\usepackage{amssymb,amsmath}
\usepackage{mathtools}
\usepackage{amsthm}
\usepackage{bbm}
\usepackage{algorithm}
\usepackage{algpseudocode}
\usepackage{accents}
\usepackage{graphicx}
\usepackage{xcolor}
\usepackage{natbib}
\usepackage{hyperref}
\usepackage{booktabs}
\usepackage{url} 
\usepackage[capitalize,noabbrev]{cleveref}
\usepackage{multirow}
\usepackage{makecell}
\usepackage{titletoc}
\usepackage{subcaption}
\newsavebox{\largestimage}
\usepackage[nodisplayskipstretch]{setspace}
\DeclareMathAlphabet\mathbfcal{OMS}{cmsy}{b}{n}

\newcommand{\blind}{0}

\theoremstyle{plain}
\newtheorem{theorem}{Theorem}[section]
\newtheorem{proposition}[theorem]{Proposition}
\newtheorem{lemma}[theorem]{Lemma}

\theoremstyle{definition}
\newtheorem{definition}[theorem]{Definition}
\newtheorem{assumption}[theorem]{Assumption}
\theoremstyle{remark}
\newtheorem{remark}[theorem]{Remark}
\newtheorem{example}[theorem]{Example}

\newcommand{\tenX}{\mathbfcal{X}}
\newcommand{\tenY}{\mathbfcal{Y}}
\newcommand{\tenW}{\mathbfcal{W}}

\newcommand{\tenE}{\mathbfcal{E}}
\newcommand{\tenB}{\mathbfcal{B}}
\newcommand{\tenT}{\mathbfcal{T}}
\newcommand{\tenF}{\mathbfcal{F}}

\newcommand{\tenA}{\mathbfcal{A}}
\newcommand{\tenC}{\mathbfcal{C}}
\newcommand{\tenG}{\mathbfcal{G}}
\newcommand{\tenV}{\mathbfcal{V}}
\newcommand{\htenX}{\widehat{\tenX}}
\newcommand{\htenB}{\widehat{\tenB}}
\newcommand{\htenT}{\widehat{\tenT}}
\newcommand{\ttenB}{\widetilde{\tenB}}
\newcommand{\ttenX}{\widetilde{\tenX}}

\newcommand{\ttenW}{\widetilde{\tenW}}
\newcommand{\trtenW}{\tenW_{\setS_{tr}}}

\newcommand{\matX}{\mathbf{X}}
\newcommand{\matY}{\mathbf{Y}}

\newcommand{\matU}{\mathbf{U}}
\newcommand{\matI}{\mathbf{I}}

\newcommand{\vecr}{\mathbf{r}}
\newcommand{\vecu}{\mathbf{u}}
\newcommand{\vecv}{\mathbf{v}}

\newcommand{\setM}{\mathbb{M}}
\newcommand{\setN}{\mathbb{N}}

\newcommand{\setS}{\mathbb{S}}
\newcommand{\setT}{\mathbb{T}}

\newcommand{\uhat}{\widehat{u}}
\newcommand{\ta}{\tilde{a}}
\newcommand{\NaN}{\text{NaN}}
\newcommand{\dk}{d_1\times\cdots\times d_K}
\newcommand{\qhat}{\widehat{q}}

\newcommand{\twonorm}[1]{\|#1\|_{\mathrm{F}}}
\newcommand{\maxnorm}[1]{\|#1\|_{\infty}}

\newcommand{\inner}[2]{\langle #1, #2\rangle}
\newcommand{\indicator}[1]{\mathbbm{1}_{\{#1\}}}

\newcommand{\proj}[2]{\mathcal{P}_{#1}(#2)}
\newcommand{\leftmat}[1]{\mathbf{L}(#1)}
\newcommand{\vect}[1]{\mathbf{vec}(#1)}
\newcommand{\Prob}{\mathrm{P}}
\newcommand{\TTSVD}[1]{\text{SVD}_{\vecr}^{\text{tt}}(#1)}

\newcommand{\ranktt}[1]{\mathrm{rank}^{\text{tt}}(#1)}

\DeclareMathOperator*{\argmin}{arg\,min}
\algnewcommand\INPUT{\item[\textbf{Input:}]}
\algnewcommand\OUTPUT{\item[\textbf{Output:}]}

\begin{document}
\def\spacingset#1{\renewcommand{\baselinestretch}%
{#1}\small\normalsize} \spacingset{1}

\if0\blind
{
  \title{\bf Conformalized Tensor Completion with Riemannian Optimization}
  \author{Hu Sun\\
    Department of Statistics, University of Michigan, Ann Arbor\\
    and \\
    Yang Chen\thanks{Email: ychenang@umich.edu
   }\hspace{.2cm} \\
    Department of Statistics and Michigan Institute for Data Science\\
    University of Michigan, Ann Arbor}
  \maketitle
} \fi

\if1\blind
{
  \bigskip
  \bigskip
  \bigskip
  \begin{center}
    {\LARGE\bf Conformalized Tensor Completion with Riemannian Optimization}
\end{center}
  \medskip
} \fi

\bigskip
\begin{abstract}
Tensor data, or multi-dimensional arrays, is a data format popular in multiple fields such as social network analysis, recommender systems, and brain imaging. It is not uncommon to observe tensor data containing missing values, and tensor completion aims at estimating the missing values given the partially observed tensor. Sufficient efforts have been spared on devising scalable tensor completion algorithms, but few on quantifying the uncertainty of the estimator. In this paper, we nest the uncertainty quantification (UQ) of tensor completion under a split conformal prediction framework and establish the connection of the UQ problem to a problem of estimating the missing propensity of each tensor entry. We model the data missingness of the tensor with a tensor Ising model parameterized by a low-rank tensor parameter. We propose to estimate the tensor parameter by maximum pseudo-likelihood estimation (MPLE) with a Riemannian gradient descent algorithm. Extensive simulation studies have been conducted to justify the validity of the resulting conformal interval. We apply our method to the regional total electron content (TEC) reconstruction problem. Supplemental materials of the paper are available online.
\end{abstract}

\noindent%
\vfill

\newpage
\spacingset{1.75} 
\section{Introduction}\label{sec:intro}
Tensor, or multi-dimensional array, has become a popular data format in several applications such as collaborative filtering~\citep{bi2018recommend}, financial time series modeling~\citep{li2021multi}, hypergraph networks analysis~\citep{ke2019community}, neuroimaging study~\citep{li2018tucker}, and astrophysics imaging analysis~\citep{sun2023tensor}. Tensor gains this popularity due to its efficient representation of structural high-dimensional data. For example, in collaborative filtering~\citep{bi2018recommend}, the rating data is naturally embedded in a 3-way tensor with each entry being the rating by a user on a certain item under a specific context. In neuroimaging analysis~\citep{wei2023tensor}, as another example, each brain voxel in the 3-way tensor is identified by its coordinate in the 3-D Euclidean space.

Tensor completion~\citep{yuan2016tensor,xia2021statistically,cai2022nonconvex} is a technique that provides an estimator of the tensor when missing values are present. Typically, given only one tensor sample with missingness, tensor completion aims at finding a low-rank tensor that best imputes the missing entries. Various optimization techniques~\citep{kressner2014low,yuan2016tensor,wang2019tensor,lee2020tensor,cai2022nonconvex,qi2023exploiting} have been proposed for computationally efficient tensor completion and the statistical error of tensor completion has also been carefully investigated~\citep{xia2021statistically}.

However, given the progress above, very little work has been done on the uncertainty quantification of tensor completion. Existing work on the uncertainty quantification of matrix completion~\citep{chen2019inference} and tensor completion~\citep{cai2022uncertainty} typically relies on asymptotic analysis of the estimator by a specific completion algorithm and assumes that data is missing uniformly at random. In this paper, we aim to devise a data-driven approach that does not rely on a specific choice of the completion algorithm nor assume the data is missing uniformly at random, which is more adaptive to real application scenarios.

Conformal prediction~\citep{vovk2005algorithmic} is a model-agnostic approach for uncertainty quantification. Recently,~\citet{gui2023conformalized} applies the idea of conformal prediction to matrix completion under the assumption that data is missing independently. The method requires one to estimate the missing propensity of each matrix entry and weigh them accordingly to construct well-calibrated confidence regions. In this paper, we generalize this idea to tensor completion. The generalization is non-trivial, as one cannot simply reshape the tensor back to a matrix for the conformal prediction without significantly increasing the dimensionality of the nuisance parameter. We keep the tensor structure and leverage low-rank tensor representations for dimension reduction. Furthermore, we do not assume data is missing independently but allow for locally dependent missingness. We capture such dependency of missingness by a novel low-rank tensor Ising model, which could be of independent interest. Finally, we propose a Riemannian gradient descent algorithm~\citep{kressner2014low} for scalable computation, which is necessary since tensor data is typically high-dimensional.

The key insight of the method is that one puts a higher weight on the tensor entries with a higher probability of missing. Such a weighted conformal prediction approach~\citep{tibshirani2019conformal} is also seen in spatial conformal prediction~\citep{mao2022valid} and localized conformal prediction~\citep{guan2023localized}, where higher weights are put on neighbors in the Euclidean or feature space. However, our method is significantly different in that we estimate the weights by using the entire tensor and determine the weights of all tensor entries altogether, while other methods determine the weight of each data locally and thus can be slow under the tensor setting.

The remainder of the paper is organized as follows. We outline the notations used in the paper in Section~\ref{subsec:notation}. Section~\ref{sec:method} describes the conformalized tensor completion (CTC) method and the probabilistic model for the data missingness. Section~\ref{sec:algorithm} is dedicated to the computational algorithm of the CTC. We validate the performance of our proposed CTC using extensive simulations in Section~\ref{sec:sim} and a real data application to a geophysics dataset in Section~\ref{sec:application}. Section~\ref{sec:conclusion} concludes. The supplemental material contains technical proofs and additional details and results of the simulation and data application.


\subsection{Notation}\label{subsec:notation}
Throughout this paper, we use calligraphic boldface letters (e.g. $\tenA,\tenB$) for tensors with at least three modes, boldface uppercase letters (e.g. $\matX,\matY$) for matrices, boldface lowercase letters (e.g. $\vecu,\vecv$) for vectors, and blackboard boldface letters (e.g. $\setS,\setT$) for sets. To index a tensor/matrix/vector, we use square brackets with subscripts such as $[\tenA]_{i_1\ldots i_K},[\matX]_{ij}, [\vecu]_i$, and will ignore the square brackets when it is clear from the context. For a positive integer $n$, we denote its index set $\{1,\ldots,n\}$ as $[n]$. For a $K$-mode tensor with size $d_1\times\cdots\times d_K$, we use $\setS$ to denote $[d_1]\times\cdots\times[d_K]$, namely the indices of all tensor entries, and we often use a single index such as $i,j,s$ instead of a $K$-tuple to denote elements from $\setS$ for notational brevity.

For any tensors $\tenX,\tenY \in \mathbb{R}^{\dk}$, we use $\vect{\tenX},\vect{\tenY}$ to denote the corresponding vectorized tensors, where all entries are aligned in such an order that the first index changes the fastest. We use $\inner{\tenX}{\tenY}$ to denote tensor inner product and basically $\inner{\tenX}{\tenY} = \vect{\tenX}^{\top}\vect{\tenY}$. Tensor Frobenius norm $\twonorm{\tenX}$ is defined as $\sqrt{\inner{\tenX}{\tenX}}$ and tensor max-norm $\maxnorm{\tenX}$ is defined as $\max_{s\in\setS} |\tenX_s|$. For any tensor $\tenX\in \mathbb{R}^{\dk}$ and any matrix $\matU\in\mathbb{R}^{J\times d_k}$, the $k$-th mode tensor-matrix product, denoted as $\tenX\times_k\matU$, is a tensor of size $d_1\times\cdots\times d_{k-1}\times J\times d_{k+1}\times\cdots \times d_k$ that satisfies:
\begin{equation*}
    [\tenX\times_k\matU]_{i_1\ldots i_{k-1}ji_{k+1}\ldots i_K} = \sum_{i_k=1}^{d_k} [\tenX]_{i_1\ldots i_k\ldots d_K}[\matU]_{ji_k}.
\end{equation*}
More preliminaries on tensor notations and the related algebra will be covered in later sections, and we refer our readers to~\citet{kolda2009tensor} for more references on the related tensor algebra. In this paper, when referring to a tensor that is a random variable, we add a tilde over the top of the tensor, such as $\ttenW,\ttenX$, and use the raw version $\tenW,\tenX$ to denote concrete samples. We add an asterisk to the superscript, such as $\tenX^*,\tenB^*$ to denote the non-random, ground truth parameters.

\section{Method}\label{sec:method}
Suppose we have a $K$-mode random tensor $\ttenX$ of size $\dk$. Further, suppose that one obtains a sample $\tenX$ for $\ttenX$ with part of the entries in $\tenX$ missing. To encode the missingness in $\tenX$, we define the binary missingness tensor $\tenW\in\{-1,1\}^{\dk}$ and set $\tenW_s=1$ when $\tenX_s$ is observed and $\tenW_s=-1$ when $\tenX_s$ is missing. We assume that the missingness $\tenW$ is a sample of a random binary tensor $\ttenW$ whose likelihood is $p(\cdot)$.

The tensor completion problem~\citep{yuan2016tensor,xia2021statistically,cai2022nonconvex} deals with estimating the values in $\tenX$ where $\tenW_s=-1$, i.e., where data is missing. Although the main framework of our paper does not rely on a specific choice of the tensor completion algorithm, it is beneficial to provide one example here, which is also the algorithm we will be using in our numerical experiments and data application.

Since one only has one sample $\tenX$ of $\ttenX$, estimating the missing values in $\tenX$ is impossible without imposing additional parsimony over the estimator. Following the literature on tensor completion~\citep{kressner2014low,xia2021statistically,cai2022provable}, we assume that the estimator has a low tensor rank and solve for the estimator by the following constrained least-square problem:
\begin{equation}\label{eq:tensor-completion-problem}
    \min_{\tenA: \mathrm{rank}(\tenA) \le \vecr} \quad \frac12 \sum_{s: \tenW_s=1} \left(\tenX_s - \tenA_s\right)^2,
\end{equation}
where the notion of tensor rank will be introduced later. We denote the minimizer of~\eqref{eq:tensor-completion-problem} as $\htenX$. The goal of the paper is to quantify the uncertainty for $\htenX$ by constructing a confidence interval $C(\htenX)$ around $\htenX$ to cover $\tenX$ with a pre-specified level of confidence. The framework, called conformalized tensor completion, will be introduced next.

\subsection{Conformalized Tensor Completion (CTC)}\label{subsec:conformal-tc}
Conformal prediction~\citep{vovk2005algorithmic} is a model-agnostic, distribution-free approach for predictive uncertainty quantification. To put in the context of the tensor completion problem, we utilize specifically the \textit{split conformal prediction}~\citep{papadopoulos2002inductive} approach for its simplicity and scalability to complex data structures such as tensor data. We leave the discussion of \textit{full conformal prediction}~\citep{shafer2008tutorial} to future work.

Split conformal prediction starts by partitioning all observed entries in $\tenX$, whose indices are denoted as $\setS_{obs}$, randomly into a training set $\setS_{tr}$ and a calibration set $\setS_{cal}$. One first provides a tensor completion estimator $\htenX$ using the training set \textit{only}, say by solving for~\eqref{eq:tensor-completion-problem} using entries in $\setS_{tr}$. Then one calculates the \textit{non-conformity score} over the calibration set by a score function $\mathcal{S}(\tenX_s,\htenX_s)$ such as $\mathcal{S}(\tenX_s,\htenX_s) = |\tenX_s-\htenX_s|$. To quantify the uncertainty of $\htenX_{s^*}$ at any missing entry $s^*\in\setS_{miss}$, where $\setS_{miss}$ includes the indices of all missing entries, the canonical conformal interval at $(1-\alpha)$ confidence level is constructed as $C_{1-\alpha,s^*}(\htenX) = \{x\in\mathbb{R}|\mathcal{S}(x,\htenX_{s^*}) \le \qhat\}$, with $\qhat$ defined as:
\begin{equation}\label{eq:qhat-definition}
    \qhat = \mathcal{Q}_{1-\alpha} \left(\frac{1}{|\setS_{cal}|+1}\cdot\sum_{s\in\setS_{cal}}\delta_{\mathcal{S}(\tenX_s,\htenX_s)} + \frac{1}{|\setS_{cal}|+1}\cdot\delta_{+\infty}\right),
\end{equation}
where $\delta_a$ is a point mass at $x=a$ and $\mathcal{Q}_{\tau}(\cdot)$ extracts the $(100\tau)^{\rm{th}}$ quantile of a distribution. The validity of such a conformal interval $C_{1-\alpha,s^*}(\htenX)$ relies on the assumption of \textit{data exchangeability}~\citep{lei2018distribution}. To put it in the context of tensor completion, we re-label $\setS_{cal}\cup\{s^*\}$ as $\{s_1,\ldots,s_{n+1}\}$, with $n = |\setS_{cal}|$ and $s_{n+1}=s^*$ and define event $\tenE_0$ as:
\begin{equation}\label{eq:event-E0-definition}
    \tenE_0=\left\{\ttenW_s=1 \text{ if and only if } s\in\setS_{tr}\cup\setS^{\prime}, \setS^{\prime}\subset\{s_1,\ldots,s_{n+1}\}\text{ and }|\setS^\prime|=n\right\}.
\end{equation}
The data exchangeability assumption is equivalent to saying that the probability:
\begin{equation*}
    \Prob\left[\ttenW_{s_k}=-1\text{ and } \ttenW_{s}=1 \text{ for } s\in\setS_k\bigg|\tenE_0\right]
\end{equation*}
is equal for all $k=1,\ldots,n+1$, where $\setS_k=\{s_1,\ldots,s_{n+1}\}\setminus\{s_k\}$. Equivalently, this states that conditioning on observing data only from $\setS_{tr}$ and $n$ out of $n+1$ entries from $\{s_1,\ldots,s_{n+1}\}$, it is equally likely to observe any $n$ entries from $\{s_1,\ldots,s_{n+1}\}$. This assumption will hold when data are missing independently with the same probability, a common assumption made in the literature on matrix/tensor completion uncertainty quantification~\citep{chen2019inference,cai2022uncertainty}. However, this assumption might not hold when the data missingness is dependent or when the missingness is independent but with heterogeneous probabilities. Therefore, it is necessary to account for more general data missing patterns when conducting uncertainty quantification.

We modify the canonical conformal prediction to accommodate more general data missing patterns by re-weighting each calibration entry using the weighted exchangeability framework~\citep{tibshirani2019conformal}. The result is summarized in Proposition~\ref{thm:conformal-interval-weight}.
\begin{proposition}\label{thm:conformal-interval-weight}
For any testing entry $s^*\in\setS_{miss}$, let $s^{*}=s_{n+1}$ and $\setS_{cal}\cup\{s^*\} = \{s_1,\ldots,s_{n+1}\}$ and $\setS_k=\{s_1,\ldots,s_{n+1}\} \setminus \{s_k\}$, then define $p_k(s^*)$ as:
\begin{equation}\label{eq:pk-definition}
    p_k(s^*) = \Prob\left(\ttenW_{s} = 1 \text{ if and only if } s\in \setS_{tr} \cup \setS_k\right),
\end{equation}
for $k=1,\ldots,n+1$. Let $\htenX$ be the output of any tensor completion method using entries only from $\setS_{tr}$ and define $\qhat_{s^*}$ as:
\begin{equation}\label{eq:qhat-weighted-definition}
    \qhat_{s^*} = \mathcal{Q}_{1-\alpha} \left(\sum_{i=1}^{n} \omega_{i}(s^*) \cdot \delta_{\mathcal{S}(\tenX_{s_i},\htenX_{s_i})} + \omega_{n+1}(s^*)\cdot\delta_{+\infty}\right),\quad \text{where }\omega_k(s^*) = \frac{p_k(s^*)}{\sum_{i=1}^{n+1} p_i(s^*)},
\end{equation}
and construct the $(1-\alpha)$-level conformal interval as $C_{1-\alpha,s^*}(\htenX) = \{x\in\mathbb{R}|\mathcal{S}(x,\htenX_{s^*}) \le \qhat_{s^*}\}$, then given the definition of $\tenE_0$ in~\eqref{eq:event-E0-definition}, we have:
\begin{equation}\label{eq:weighted-conformal-coverage-guarantee}
    \Prob\left(\tenX_{s^*} \in C_{1-\alpha,s^*}(\htenX)\bigg|\tenE_0\right) \ge 1-\alpha.
\end{equation}
\end{proposition}
We provide the detailed proof in Appendix~\ref{app:proof-weighted-conformal-coverage}. Proposition~\ref{thm:conformal-interval-weight} indicates that as long as one can properly weight each calibration entry in proportion to $p_k(s^*)$ as defined in~\eqref{eq:pk-definition}, one can obtain the conditional coverage guarantee in~\eqref{eq:weighted-conformal-coverage-guarantee}. A similar result to Proposition~\ref{thm:conformal-interval-weight} has been established for conformalized matrix completion~\citep{gui2023conformalized}, where the data is assumed to be missing independently. In our paper, we do not assume independent missingness but provide a more general statement that requires one to weight each calibration and testing entry by directly evaluating the likelihood of $\ttenW$ under $n+1$ different missingness, where each time we set $1$ out of $n+1$ entries as missing. In Section~\ref{subsec:missing-prop-model}, we will formally introduce the likelihood of the binary tensor $\ttenW$ that nests the independent missingness as a special case.




\subsection{Missing Propensity Model}\label{subsec:missing-prop-model}
The key to constructing the conformal interval with coverage guarantee is to properly weight each calibration sample by $p_k(s^*)$ in~\eqref{eq:pk-definition}, which requires the knowledge of the likelihood of $\ttenW$. In practice, one does not have access to such knowledge but needs to estimate the likelihood of $\ttenW$, given a single sample $\tenW$, and then plug in~\eqref{eq:pk-definition} to get an estimator $\widehat{p}_k(s^*)$. Previous works~\citep{chen2019inference,cai2022uncertainty,gui2023conformalized} assume that all matrix/tensor entries are missing independently, potentially with heterogeneous probabilities. This assumption, however, is not general enough. For example, for spatio-temporal tensors, data might be missing together if located close in space or time. 

Accounting for the dependencies of binary random variables turns out to be even more challenging in our context because all the binary random variables in $\ttenW$ are embedded in a tensor grid with ultra-high dimensionality. In this paper, we do not account for arbitrary data missing patterns but focus on independent missingness and locally dependent missingness. These two types of missingness are common for many data applications, such as recommender systems~\citep{bi2018recommend}, neuro-imaging~\citep{li2017low}, and remote sensing~\citep{sun2022matrix}. A more flexible dependency structure for missingness could be modeled; however, for tractability, we limit our focus to these two settings via the lens of the Ising model~\citep{cipra1987introduction}, which provides a way of modeling dependency among binary random variables. 


To start with, the Ising model prescribes a Boltzmann distribution for $\ttenW$: $p(\ttenW) \propto \exp[-\beta\mathcal{H}(\ttenW)]$, where $\beta > 0$ is the inverse temperature parameter and $\mathcal{H}(\ttenW)$ is the \textit{Hamiltonian} of $\ttenW$, describing the ``energy" of $\ttenW$. In our paper, we extend the richness of this model by augmenting $p(\ttenW)$ with an unknown tensor parameter $\tenB\in\mathbb{R}^{\dk}$ such that:
\begin{equation}\label{eq:boltzmann-distribution}
    p(\ttenW|\tenB) \propto \exp\{-\mathcal{H}(\ttenW|\tenB)\}
\end{equation}
\begin{equation}\label{eq:hamiltonian}
    \mathcal{H}(\ttenW|\tenB) = -\frac12\sum_{i\sim j} g(\tenB_i,\tenB_j)\ttenW_i\ttenW_j - \sum_i h(\tenB_i)\ttenW_i,
\end{equation}
where $i,j\in[d_1]\times\cdots\times[d_K]$, $g(\cdot,\cdot)$ and $h(\cdot)$ being pre-specified functions with $\beta$ incorporated, and $i\sim j$ means that the two entries indexed by $i$ and $j$ are ``neighbors". For brevity, we often denote $g(\tenB_i,\tenB_j)$ as $g_{ij}$ and $h(\tenB_i)$ as $h_i$ for any $i,j$. We call~\eqref{eq:boltzmann-distribution} and~\eqref{eq:hamiltonian} our missing propensity model. 

One can interpret the unknown parameter $\tenB$ as a 1-dimensional feature of each tensor entry. Each neighboring pair of entries $i$ and $j$ contribute to the Hamiltonian via their ``co-missingness" $\ttenW_i\ttenW_j$ and the interaction of their features $\tenB_i,\tenB_j$ through $g(\tenB_i,\tenB_j)$. The function $g(\cdot,\cdot)$ describes the tendency of neighboring entries to be observed or missing together. Every entry $i$ also contributes to the Hamiltonian via $h(\tenB_i)$, where the function $h(\cdot)$ describes the tendency of each entry to be observed or missing. Here we provide two concrete examples of the model.

\begin{example}[Bernoulli Model]\label{example:Bernoulli-model}
Suppose that $g(\cdot,\cdot)=0$, and let $h(x) = 0.5\cdot\log f(x)/[1-f(x)]$, where $f(\cdot)$ is an inverse link function (e.g., sigmoid function). The missing propensity model indicates that every $s\in[d_1]\times\cdots\times[d_K]$ is missing independently with:
\begin{equation}\label{eq:Bernoulli-model}
    \ttenW_s = \begin{cases}
        1, & p = f(\tenB_s) \\
        -1, & p = 1 - f(\tenB_s).
    \end{cases}
\end{equation}
\end{example}

\begin{example}[Ising Model]\label{example:Ising-model}
Suppose that $h(\cdot)=x/2$, and let $g(x,y)=xy$. Under this scenario, the conditional distribution of $\ttenW_s$, given all other entries in $\ttenW$ as $\ttenW_{-s}$, is:
\begin{equation}\label{eq:Ising-model-no-ext}
    p(\ttenW_s=1|\tenB,\ttenW_{-s}) = \frac{\exp\left[2\tenB_s\sum_{j\in\mathcal{N}(s)}\ttenW_j\tenB_j + \tenB_s\right]}{1 + \exp\left[2\tenB_s\sum_{j\in\mathcal{N}(s)}\ttenW_j\tenB_j + \tenB_s\right]} = f(\tenB_s|\sigma_s),
\end{equation}
where $\mathcal{N}(s) = \{j\in[d_1]\times\cdots\times[d_K]|s\sim j\}$, and $f(x|\sigma) = [1+\exp(-x/\sigma)]^{-1}$ is the sigmoid function with scale parameter $\sigma$. This model is similar to the Bernoulli model in~\eqref{eq:Bernoulli-model} but has an entry-specific scale parameter $\sigma_s = (2\sum_{j\in\mathcal{N}(s)}\ttenW_j\tenB_j+1)^{-1}$ that depends on the missingness and feature of the neighboring entries. 
\end{example}

Our missing propensity model shares several similarities with the previous literature on modeling the missingness of matrix/tensor data. \citet{liang2016modeling, schnabel2016recommendations, wang2018collaborative} model the missing probability via a logistic regression model with mode-specific features, which is similar to our setup in Example~\ref{example:Bernoulli-model} with $\tenB$ having low CP rank. \citep{ma2019missing} models the missing probability via denoising the binary missingness mask with a nuclear-norm penalty~\citet{davenport20141}, which is similar to our low-rank setting introduced later. Our model is distinct in the sense that it explicitly models the local dependency of the missingness, as characterized by the neighboring structure and the bivariate function $g(\cdot,\cdot)$.

Given the missing propensity model in~\eqref{eq:boltzmann-distribution} and~\eqref{eq:hamiltonian}, we can compute the $p_k(s^*)$ according to~\eqref{eq:pk-definition} and obtain the conformal weight $\omega_k(s^*)$ as:
\begin{equation}\label{eq:Ising-model-conformal-weight-exact}
\omega_k(s^*) = \frac{p_k(s^*)}{\sum_{i=1}^{n+1} p_i(s^*)} = \frac{\exp\left[-2\sum_{s_j\in\mathcal{N}(s_k)} g(\tenB_{s_k},\tenB_{s_j})\ttenW_{s_j}-2h(\tenB_{s_k})\right]}{\sum_{i=1}^{n+1} \exp\left[-2\sum_{s_j\in\mathcal{N}(s_i)} g(\tenB_{s_i},\tenB_{s_j})\ttenW_{s_j}-2h(\tenB_{s_i})\right]},
\end{equation}
with $s_1,\ldots,s_n\in\setS_{cal}, s^*=s_{n+1}$, and $\ttenW_s=1$ only if $s\in\setS_{tr}\cup\setS_{cal}\cup\{s^*\}$. The dependency of $\omega_k(s^*)$ on $s^*$ makes it computationally inefficient to scale~\eqref{eq:Ising-model-conformal-weight-exact} to all $s^*\in\setS_{miss}$ since one has to temporarily set $\ttenW_{s^*}=1$ to compute all the weights. To speed up the computation, we approximate the weight in~\eqref{eq:Ising-model-conformal-weight-exact} by plugging in $\ttenW_{s} = \tenW_s$ for all $s$, which removes the dependency of $\omega_k$ on $s^*$. Although this simplification means that one cannot obtain the exact conformal weights, in Appendix~\ref{app:weight-error-bound}, we show that the error caused by this approximation over the distribution of the non-conformity score in~\eqref{eq:qhat-weighted-definition} is negligible, and we also show this empirically in Section~\ref{sec:sim}.

With this approximation, the conformal weight $\omega_k$ is now proportional to $(1-\tilde{p}_{s_k}) / \tilde{p}_{s_k}$, where $\tilde{p}_s = p(\ttenW_{s}=1|[\ttenW]_{s^{\prime}} = [\tenW]_{s^{\prime}}, \forall s^{\prime}\neq s)$ is the full conditional probability of entry $s$ being observed given all other entries. Next, we will discuss the estimation of $\tenB$ given $\tenW$.




\section{Estimating Algorithm}\label{sec:algorithm}
In this section, we discuss the details of estimating $\tenB$ based on a single binary tensor sample $\tenW$ drawn from the missing propensity model specified by~\eqref{eq:boltzmann-distribution} and~\eqref{eq:hamiltonian}. More specifically, we attempt to estimate $\tenB$ using $\tenW_{\setS_{tr}}$, the binary tensor with $\tenW_{s}=1$ if and only if $s\in\setS_{tr}$. We describe the estimation framework in Section~\ref{subsec:MPLE-framework} and the algorithm in Section~\ref{subsec:RGrad-algorithm}.


\subsection{Low-rank MPLE Framework}\label{subsec:MPLE-framework}
Since we only have access to one sample $\tenW$ and the tensor parameter $\tenB$ is of the same dimensionality as $\tenW$, it is infeasible to obtain an estimator $\htenB$ without imposing additional constraints over $\tenB$. Similar to previous literature~\citep{wang2020learning,cai2022generalized}, we assume that the tensor $\tenB$ has low tensor rank.

In this paper, we assume that the tensor $\tenB$ has a low Tensor-Train (TT) rank~\citep{oseledets2011tensor}. A low TT-rank tensor $\tenA$ can be represented by a series of 3-mode TT factor tensors $\tenF_k\in\mathbb{R}^{r_{k-1}\times d_k\times r_k}, k=1,\ldots, K, r_0=r_K=1$, where for every entry of $\tenA$, one has:
\begin{equation}\label{eq:TT-format}
    [\tenA]_{i_1,\ldots,i_K} = [\tenF_1]_{:i_1:} [\tenF_2]_{:i_2:} \cdots [\tenF_K]_{:i_K:}, 
\end{equation}
with the right-hand side being a series of matrix multiplications. We say $\vecr = (r_1,\ldots,r_{K-1})$ is the TT-rank of $\tenA$ and compactly, we write $\tenA = [\tenF_1,\ldots,\tenF_K]$ and $\ranktt{\tenA}=\vecr$. As compared to the more commonly used Tucker rank~\citep{kolda2009tensor}, the Tensor-Train rank ensures that the number of parameters representing a low-rank tensor scales linearly with $K$, the number of modes, making the low TT-rank tensors more efficient for representing high-order tensors.

To ensure the identifiability of TT factors $\tenF_1,\ldots,\tenF_K$ in~\eqref{eq:TT-format}, it is often required that $\tenF_{1},\ldots,\tenF_{K-1}$ being \textit{left-orthogonal}. A 3-mode tensor $\tenF \in \mathbb{R}^{d_1\times d_2\times d_3}$ is left-orthogonal if $\leftmat{\tenF}^{\top}\leftmat{\tenF} = \matI_{d_3\times d_3}$, where $\leftmat{\cdot}: \mathbb{R}^{d_1\times d_2\times d_3} \mapsto \mathbb{R}^{(d_1d_2)\times d_3}$ is the so-called left-unfolding operator. Finding the representation~\eqref{eq:TT-format} of a low TT-rank tensor under the left orthogonality constraint can be achieved by the TT-SVD algorithm~\citep{oseledets2011tensor}. For completeness, we restate the TT-SVD algorithm in Algorithm~\ref{alg:TT-SVD} and denote it as $\TTSVD{\cdot}$.

\begin{algorithm}[!thb]
\caption{Tensor-Train Singular Value Decomposition (TT-SVD)}\label{alg:TT-SVD}
\begin{algorithmic}
    \INPUT Tensor $\tenX\in\mathbb{R}^{\dk}$, tensor-train rank $\vecr=(r_1,\ldots,r_{K-1})$.
    \State $\tenA \leftarrow \tenX$, $r_0,r_K\leftarrow 1$.
    \For{$k=1,\dots,K-1$}
        \State $\mathbf{A}\leftarrow \text{reshape}[\tenA, (r_{k-1}d_k,d_{k+1}\cdots d_K)]$. \hfill \% \text{reshape}$(\cdot,\cdot)$ from MATLAB 
        \State Conduct SVD on $\mathbf{A}$ and truncate at rank $r_{k}$: $\mathbf{A} \approx \matU\mathbf{S}\mathbf{V}^{\top}$.
        \State $\tenF_k \leftarrow \text{reshape}[\matU, (r_{k-1},d_k,r_k)]$.
        \State $\tenA \leftarrow \mathbf{S}\mathbf{V}^{\top}$.
    \EndFor
    \State $\tenF_K \leftarrow \text{reshape}(\tenA, (r_{K-1},d_K,1))$.
    \OUTPUT Tensor-Train representation $\htenX = [\tenF_1,\ldots,\tenF_K]$ with $\text{rank}^{\text{tt}}(\htenX) \le \vecr$.
\end{algorithmic}
\end{algorithm}

Given the assumption that the tensor $\tenB$ has low TT-rank $\vecr = (r_1,\ldots,r_{K-1})$, we can re-formulate the MLE of $\tenB$ as the solution of a low-rank tensor learning problem:
\begin{equation}\label{eq:low-TT-rank-learning-problem}
    \htenB = \argmin_{\tenB: \ranktt{\tenB} \le \vecr} -\log p(\ttenW = \trtenW|\tenB),
\end{equation}
where $\ranktt{\tenB} = (r_1^\prime,\ldots,r_{K-1}^{\prime}) \le \vecr$ means that $r_k^{\prime} \le r_k$ for any $k=1,\ldots,K-1$.

However, the likelihood in~\eqref{eq:low-TT-rank-learning-problem} is incorrect since we did not account for the random splitting of the training set and the calibration set, and it is also difficult to evaluate the normalizing constant of the likelihood. To tackle these issues, we consider estimating $\tenB$ by the maximum pseudo-likelihood estimator (MPLE), which is a common approach for the estimation and inference of the Ising model~\citep{ravikumar2010high,barber2015high,bhattacharya2018inference}. Formally, for each entry $i$, define $\tilde{p}_i(\tenB)$ as:
\begin{align}
    \tilde{p}_i(\tenB) & = p\left(\ttenW_i=1|[\ttenW]_s=[\tenW_{\setS_{tr}}]_s, \forall s\neq i, \tenB\right) \nonumber \\
    & = \frac{\exp\left[2\sum_{j\in\mathcal{N}(i)}g(\tenB_i,\tenB_j)[\trtenW]_j + 2h(\tenB_i)\right]}{1 + \exp\left[2\sum_{j\in\mathcal{N}(i)}g(\tenB_i,\tenB_j)[\trtenW]_j + 2h(\tenB_i)\right]}. \label{eq:pseudo-pi}
\end{align}
and we often write it directly as $\tilde{p}_i$. The low-rank MPLE of $\tenB$ can now be written as:
\begin{equation}\label{eq:low-rank-MPLE}
    \htenB = \argmin_{\tenB: \ranktt{\tenB} \le \vecr} \ell(\trtenW|\tenB) = -\sum_{i: [\tenW_{\setS_{tr}}]_i=1} \log q\tilde{p}_i - \sum_{i: [\tenW_{\setS_{tr}}]_i=-1} \log\left(1-q\tilde{p}_i\right),
\end{equation}
where $q\in(0,1)$ is the probability of selecting an observed entry into the training set. We discuss the optimization algorithm for solving~\eqref{eq:low-rank-MPLE} next.


\subsection{Riemannian Gradient Descent (RGrad) Algorithm}\label{subsec:RGrad-algorithm}
To solve for~\eqref{eq:low-rank-MPLE}, a natural idea is to directly estimate the tensor-train factors $\tenT_1,\ldots,\tenT_K$ for $\htenB$ one at a time, while keeping the others fixed, and iterate until convergence. Such an alternating minimization algorithm has been applied to low-rank binary tensor decomposition~\citep{wang2020learning,lee2020tensor}. However, alternating minimization is computationally inefficient here as each step requires fitting a generalized linear model (GLM) with high-dimensional covariates. Another candidate approach for estimating $\htenB$ is the projected gradient descent~\citep{chen2019non}, where in each iteration one updates $\tenB$ along the gradient direction first and then projects it back to the low-rank tensor space with TT-SVD. This is also undesirable since the projection for a high-rank tensor can be very slow.

In this paper, we propose an optimization technique called Riemannian gradient descent (RGrad), motivated by the fact that rank-$\vecr$ tensor-train tensors lie on a smooth manifold~\citep{holtz2012manifolds}, which we denote as $\setM_{\vecr}$. As compared to the aforementioned methods, RGrad is faster because each step updates $\tenB$ with a gradient along the tangent space of $\tenB$, avoiding fitting multiple high-dimensional GLMs. Also, the projection from the tangent space back to the manifold $\setM_{\vecr}$ is faster than the projected gradient descent since the tensors in the tangent space are also low-rank. RGrad has been extensively applied to tensor completion~\citep{kressner2014low,steinlechner2016riemannian,cai2022provable}, generalized tensor learning~\citep{cai2022generalized} and tensor regression~\citep{luo2022tensor}. The current work, to the best of our knowledge, is the first to apply RGrad to the low TT-rank binary tensor decomposition. We break down the procedures of RGrad into three steps.


\textit{\underline{Step I: Compute Vanilla Gradient.}} We first compute the vanilla gradient $\nabla\ell(\trtenW|\tenB)$ at the current iterative value $\tenB$. Formally, the vanilla gradient tensor $\tenG$ satisfies:
\begin{equation}\label{eq:vanilla-gradient}
    [\tenG]_{i} = 2\sum_{j\in\mathcal{N}(i)} \left(\tenV_i[\trtenW]_j + \tenV_j[\trtenW]_i\right) g_x(\tenB_i,\tenB_j) + 2h^{\prime}(\tenB_i)\tenV_i,
\end{equation}
where $g_x(\cdot,\cdot) = \partial g(\cdot,\cdot) / \partial x$ and $\tenV_i = (1-\tilde{p}_i)(1-q\tilde{p}_i)^{-1}(q\tilde{p}_i - \indicator{[\trtenW]_i=1})$, with $\tilde{p}_i$ defined in~\eqref{eq:pseudo-pi}. Typically, we require the neighboring structure $\mathcal{N}(i)$ to be symmetric across all entries $i$ and require function $g(x,y)$ to be a bivariate polynomial of the form $\sum_{\alpha,\beta}c_{\alpha,\beta}x^\alpha y^\beta$, which would then allow one to compute~\eqref{eq:vanilla-gradient} much faster with convolutions.


\textit{\underline{Step II: Tangent Space Projected Gradient Descent.}} 
Suppose that the current iterative value $\tenB$ has a tensor-train representation $\tenB = [\tenT_1,\ldots,\tenT_K]$. Then any tensor $\tenA$ within the tangent space $\setT$ at $\tenB$ has an explicit form:
\begin{equation}\label{eq:tangent-space-tensor-form}
    \tenA = \sum_{k=1}^{K} \tenC_k, \quad \tenC_k = [\tenT_1,\ldots,\tenT_{k-1},\tenY_k,\tenT_{k+1},\ldots,\tenT_K],
\end{equation}
with the constraint that $\leftmat{\tenY_k}^{\top}\leftmat{\tenT_k} = \mathbf{O}_{r_k\times r_k}$ for all $k < K$, where $\mathbf{O}$ is a zero matrix, and $\tenC_k$ has the property that $\inner{\tenC_i}{\tenC_j} = 0$ for all $i\neq j$. In this step, one projects the vanilla gradient $\tenG$ from step I onto $\setT$ and obtains the projected gradient $\proj{\setT}{\tenG}$. Thanks to the orthogonality of different $\tenC_k$, the projection problem can be solved via:
\begin{equation}\label{eq:tangent-space-projection-problem}
    \min_{\tenY_k: \leftmat{\tenY_k}^{\top}\leftmat{\tenT_k} = \mathbf{O}_{r_k\times r_k}} \frac12 \twonorm{\tenG - \tenC_k}^2, \quad \text{s.t. } \tenC_k = [\tenT_1,\ldots,\tenT_{k-1},\tenY_k,\tenT_{k+1},\ldots,\tenT_K],
\end{equation}
for any $k\le K-1$ and $\tenY_k$ is unconstrained if $k=K$. Solution to~\eqref{eq:tangent-space-projection-problem} is:
\begin{equation}\label{eq:tangent-space-projection-solution-1}
    \leftmat{\widehat{\tenY}_k} = \left[\matI_{r_{k-1}d_k} - \leftmat{\tenT_k}\leftmat{\tenT_k}^{\top}\right]\left(\tenB^{\le k-1}\otimes \matI_{d_k}\right)^{\top}\tenG^{<k>}\left(\tenB^{\ge k+1}\right)^{\top}\left[\tenB^{\ge k+1}\left(\tenB^{\ge k+1}\right)^{\top}\right]^{-1},
\end{equation}
for $k\le K-1$ and:
\begin{equation}\label{eq:tangent-space-projection-solution-2}
    \leftmat{\widehat{\tenY}_K} = \left(\tenB^{\le K-1}\otimes \matI_{d_K}\right)^{\top}\tenG^{<K>},
\end{equation}
where $\otimes$ is the matrix Kronecker product. In~\eqref{eq:tangent-space-projection-solution-1} and~\eqref{eq:tangent-space-projection-solution-2}, $\tenG^{<k>}$ is the $k$-mode separation of tensor $\tenG$, which basically reshapes $\tenG$ to a matrix of size $\left(\prod_{l\le k} d_l\right) \times \left(\prod_{l > k} d_l\right)$. Any tensor $\tenB$ has its $k$-mode separation as $\tenB^{<k>} = \tenB^{\le k}\tenB^{\ge k+1}$, where $\tenB^{\le k},\tenB^{\ge k+1}$ are called the $k$-th left part and $(k+1)$-th right part. Given that $\tenB = [\tenT_1,\ldots,\tenT_K]$, one can recursively compute $\tenB^{\le k}$ as $(\tenB^{\le k-1}\otimes \matI_{d_k})\leftmat{\tenT_k}$ and $\tenB^{\ge k+1}$ as $\mathbf{R}(\tenT_{k+1})(\matI_{d_{k+1}}\otimes \tenB^{\ge k+2})$ following the convention that $\tenB^{\le 0}=\tenB^{\ge K+1}=1$, where $\mathbf{R}(\cdot): \mathbb{R}^{d_1\times d_2\times d_3}\mapsto \mathbb{R}^{d_1\times d_2d_3}$ is the right-unfolding operator. 

After computing $\widehat{\tenY}_k$ with~\eqref{eq:tangent-space-projection-solution-1} and~\eqref{eq:tangent-space-projection-solution-2}, one ends up with $\widehat{\tenC}_k = [\tenT_1,\ldots,\widehat{\tenY}_k,\ldots,\tenT_K]$ and thus the projected gradient $\proj{\setT}{\tenG} = \sum_{k}\widehat{\tenC}_k$. With this projected gradient, one updates $\tenB$ via $\widetilde{\tenB} \leftarrow \tenB - \eta\proj{\setT}{\tenG}$, where $\eta$ is a constant step size.

\textit{\underline{Step III: Retraction.}} As a property of low TT-rank tensors, the updated tensor $\ttenB$ has its TT-rank upper bounded by $2\vecr$. To enforce the rank constraint, the last step of RGrad is to retract $\ttenB$ back to the manifold $\setM_{\vecr}$. We do so by applying TT-SVD to $\ttenB$: $\tenB^{\prime} = \TTSVD{\ttenB}$, and $\tenB^{\prime}$ will be the value used for the next iteration.

We summarize the RGrad algorithm in Algorithm~\ref{alg:RGrad}. To provide an initial estimator of $\tenB$, we apply TT-SVD to a randomly perturbed version of the binary tensor $\trtenW$, which works quite well empirically. We typically set $\eta = 0.1$ and denote the output of Algorithm~\ref{alg:RGrad} as $\text{RGrad}(\trtenW,\vecr)$. The algorithm terminates when $\twonorm{\tenB^\prime - \tenB}$ falls below a prespecified threshold.
\begin{algorithm}
\caption{MPLE of Low-rank Ising Model with Riemannian Gradient Descent}\label{alg:RGrad}
\begin{algorithmic}
\INPUT Binary tensor $\trtenW$, tensor-train rank $\vecr=(r_1,\ldots,r_{K-1})$, step size $\eta$, train-calibration split probability $q$.
\State Initialize: let $\tenE \overset{i.i.d.}{\sim} \mathcal{N}(0,\sigma^2)$ and $\htenB \leftarrow \TTSVD{\trtenW + \tenE}=[\htenT_1,\ldots,\htenT_K]$ by Algorithm~\ref{alg:TT-SVD}.
\For{$l=1,\ldots,l_{\text{max}}$}
\State Compute the vanilla gradient $\tenG$ using~\eqref{eq:vanilla-gradient}.
\For{$k=1,\ldots,K$}
\State Compute $\widehat{\tenY}_k$ following~\eqref{eq:tangent-space-projection-solution-1} if $k < K$ and~\eqref{eq:tangent-space-projection-solution-2} if $k=K$.
\State $\widehat{\tenC}_k \leftarrow [\htenT_1,\ldots,\htenT_{k-1},\widehat{\tenY}_k,\htenT_{k+1},\ldots,\htenT_K]$.
\EndFor
\State $\proj{\setT}{\tenG} \leftarrow \sum_{k=1}^K \widehat{\tenC}_k$.
\State $\ttenB \leftarrow \htenB - \eta\proj{\setT}{\tenG}$.
\State $\htenB \leftarrow \TTSVD{\ttenB} = [\htenT_1,\ldots,\htenT_K]$ by Algorithm~\ref{alg:TT-SVD}.
\EndFor
\OUTPUT Maximum Pseudo-Likelihood Estimator (MPLE) $\htenB$ with $\ranktt{\htenB} \le \vecr$.
\end{algorithmic}
\end{algorithm}
By assuming $d_k = O(d), r_k = O(r), \forall k$ and $\max_s |\mathcal{N}(s)| = O(K)$, the computational complexity of RGrad is $O(K(d^Kr^2+dr^3))$ per iteration. See~\citet{steinlechner2016riemannian} for more details on the computational complexity of RGrad. In Appendix~\ref{app:time_n_conv}, we verify the numerical convergence of Algorithm~\ref{alg:RGrad}, making runtime comparisons with competing methods and exploring an adaptive stepsize scheme. We summarize the conformalized tensor completion (CTC) algorithm in Algorithm~\ref{alg:CTC}.
\begin{algorithm}
\caption{Conformalized Tensor Completion (CTC)}\label{alg:CTC}  
\begin{algorithmic}
\INPUT Data tensor $\tenX$, tensor-train rank $\vecr$, train-calibration split probability $q\in(0,1)$, target mis-coverage $\alpha\in(0,1)$, arbitrary tensor completion algorithm $\tenA$.
\State $\setS \leftarrow \{s\in[d_1]\times\cdots\times[d_K]|\tenX_s\neq \NaN\}$. \hfill \% \text{indices of entries that are observed}
\State $\tenW \leftarrow 2\times \indicator{s\in\setS} - 1$.
\State Randomly partition $\setS$ independently into $\setS_{tr}\cup\setS_{cal}$ with probability $q$ and $1-q$.
\State $\htenX \leftarrow \tenA(\tenX_{\setS_{tr}})$. \hfill \% $[\tenX_{\setS_{tr}}] = \tenX_s$ if $s\in\setS_{tr}$ and $\NaN$ otherwise
\State $\htenB \leftarrow \text{RGrad}(\trtenW,\vecr)$. \hfill \% $\text{RGrad}(\cdot,\cdot)$ is Algorithm~\ref{alg:RGrad}
\For{$s\in\setS_{cal}\cup\setS^{c}$}
\State $\tilde{p}_s \leftarrow \left\{1 + \exp\left[-2\sum_{j\in\mathcal{N}(s) }[\trtenW]_jg(\htenB_s,\htenB_j) - 2h(\htenB_s)\right]\right\}^{-1}$.
\State $\omega_{s} \leftarrow (1-\tilde{p}_s)\tilde{p}_s^{-1}$. 
\EndFor
\For{$s^*\in\setS^{c}$} \hfill \% \text{See Remark~\ref{rmk:test-conformal-weight}}
\State Re-normalize $\omega_s,s\in\setS_{cal}$ and $\omega_{s^*}$ s.t. $\sum_{s\in\setS_{cal}}\omega_s + \omega_{s^*} = 1$.
\State $\qhat_{s^*} \leftarrow \mathcal{Q}_{1-\alpha} \left(\sum_{s\in\setS_{cal}} \omega_{s} \cdot \delta_{\mathcal{S}(\tenX_s,\htenX_s)} + \omega_{s^*}\cdot\delta_{+\infty}\right).$
\EndFor
\OUTPUT $(1-\alpha)$-level conformal interval $C_{1-\alpha,s^*}(\htenX) \leftarrow \{x\in\mathbb{R}|\mathcal{S}(x,\htenX_{s^*}) \le \qhat_{s^*}\}, \forall s^*\in\setS^c$.
\end{algorithmic}
\end{algorithm}
\begin{remark}[Fast Entry-wise Quantile Computation]\label{rmk:test-conformal-weight}
In the last step of Algorithm~\ref{alg:CTC}, we compute the empirical $(1-\alpha)$-quantile of the weighted eCDF of the non-conformity score of all calibration data. The for-loop looks slow, as one needs to evaluate the quantile for each testing entry $s^*$ separately. However, $\qhat_{s^*}$ can be computed faster via:
\begin{equation*}
    \qhat_{s^*} = \begin{cases}
    +\infty, & \text{if } \omega_{s^*} \ge \alpha \\
    \mathcal{Q}_{\frac{1-\alpha}{1-\omega_{s^*}}} \left(\sum_{s\in\setS_{cal}} \frac{\omega_{s}}{1-\omega_{s^*}} \cdot \delta_{\mathcal{S}(\tenX_s,\htenX_s)}\right), & \text{if } \omega_{s^*} < \alpha.
    \end{cases}
\end{equation*}
which only requires one to evaluate the quantile of a fixed weighted eCDF shared by all testing entries, and can be computed efficiently with sorting-based approaches or quantile sketching~\citep{greenwald2001space} if the calibration set is large in size.
\end{remark}
\begin{remark}[Rank Selection]\label{rmk:model-selection}
The implementation of the CTC algorithm requires a proper choice of the tensor-train rank $\vecr$ for the low-rank Ising model. Typically, in low-rank tensor learning literature~\citep{wang2020learning,cai2022generalized}, either the Akaike Information Criterion (AIC)~\citep{akaike1973information} or the Bayesian Information Criterion (BIC)~\citep{schwarz1978estimating} is used for the rank selection. Unfortunately, they are not applicable here since we can only compute the pseudo-likelihood. According to previous literature on the model selection of Markov Random Fields~\citep{ji1996consistent,csiszar2006consistent,matsuda2021information}, one can replace the likelihood in AIC/BIC with pseudo-likelihood and obtain the Pseudo-AIC (P-AIC) and Pseudo-BIC (P-BIC), which are still consistent under some regularity conditions. The P-AIC and P-BIC are defined as:
\begin{equation}\label{eq:P-AIC-criterion}
    \text{P-AIC}(r^{\prime}) = 2\ell(\trtenW|\htenB) + 2\left\{\sum_{k=1}^{K-1} \left[d_kr^{\prime}_{k-1}r^{\prime}_k-(r^{\prime}_k)^2\right] + d_{K}r_{K-1}^{\prime}\right\}.
\end{equation}
\begin{equation}\label{eq:P-BIC-criterion}
    \text{P-BIC}(r^{\prime}) = 2\ell(\trtenW|\htenB) + \left\{\sum_{k=1}^{K-1} \left[d_kr^{\prime}_{k-1}r^{\prime}_k-(r^{\prime}_k)^2\right] + d_{K}r_{K-1}^{\prime}\right\} \log\left(\prod_{k=1}^{K} d_k\right).
\end{equation}
Among all candidate ranks, we select the rank with the smallest P-AIC or P-BIC. In Section~\ref{app:decomp-simulation} of the supplemental material, we provide empirical evidence on the consistency of P-AIC and the inconsistency of P-BIC.
\end{remark}
\begin{remark}[Estimation and Coverage Error Bound]\label{rmk:theoretical-error-bound}
In Section~\ref{app:error-bound} of the supplemental material, we derive theoretically the non-asymptotic bound for $\twonorm{\htenB-\tenB^*}$ under the special case where $g(x,y)=0$ (i.e. the Bernoulli model) and with the same assumption we further derive the coverage probability lower bound of the CTC algorithm in Section~\ref{app:cov-guarantee}. It is a remarkable result that the estimating error, as well as the shortfall of the coverage from the target coverage, increases with $(r^*\bar{d}/d^*)^{1/2}$, where $r^*,\bar{d},d^*$ are $\prod_k r_k, \sum_k d_k, \prod_k d_k$ for $\tenB^*$, respectively. If one assumes that $d_k=O(d), r_k=O(r), \forall k$, then the estimation error and coverage shortfall scales with $(r/d)^{(K-1)/2}$. Higher $r/d$ indicates that the data missing pattern is more complex and thus the uncertainty quantification is harder. However, we do not have the results when $g(x,y)\neq 0$ given theoretical challenges, we show empirically in Section~\ref{sec:sim} that this tendency also holds for the Ising model.
\end{remark}


\section{Simulation Experiments}\label{sec:sim}
In this section, we validate the effectiveness of the proposed conformalized tensor completion algorithm via numerical simulations. We consider an order-3 cubical tensor of size $d\times d\times d$ and summarize our simulation settings below. Additional details about the simulation setups and results are included in Section~\ref{app:simulation} of the supplemental material. Our code is available on \href{https://github.com/husun0822/ConformalTensor}{GitHub}.

\subsection{Simulation Setup}\label{subsec:sim-setup}
\begin{figure}[!htb]
    \centering
    \includegraphics[width=0.98\textwidth]{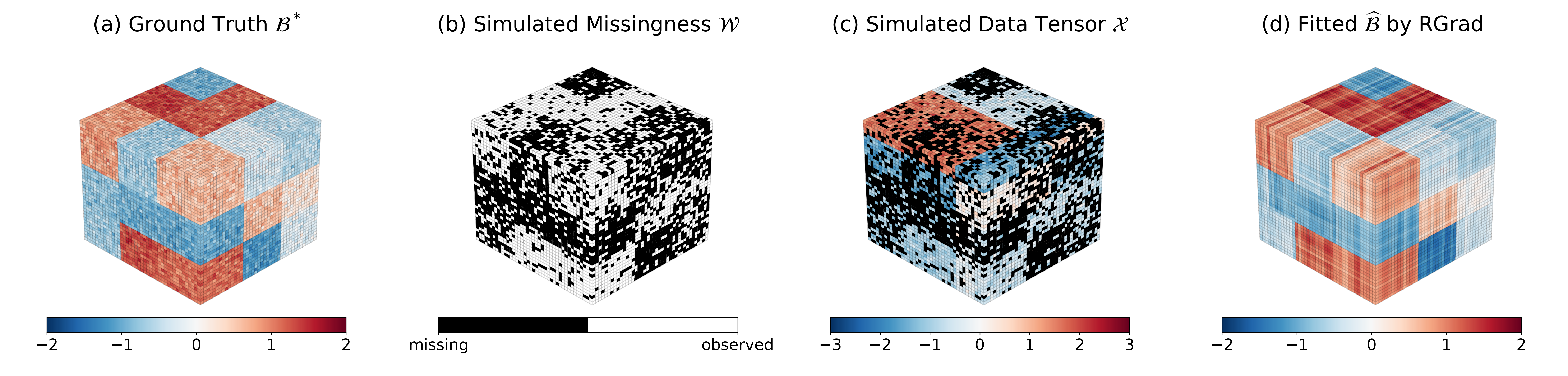}
    \caption{Visualizations of key tensors in the simulation setup. (a) Ising model parameter tensor $\tenB^*$ with $d=40,r=3$. (b) Simulated binary tensor $\tenW$ with $g(x,y)=xy/15, h(x)=x/2$. (c) Simulated data tensor $\tenX$ masked by $\tenW$ with $r_0=3, \text{SNR}=2.0$ and $\tenE$ having i.i.d. $\mathcal{N}(0,1)$ entries. (d) Estimated parameter $\htenB$ from RGrad based on a $70\%$ training set.}
    \label{fig:simulation-setup}
\end{figure}

We simulate the $d\times d\times d$ true tensor parameter $\tenB^*$ via the Gaussian tensor block model (TBM)~\citep{wang2019multiway}, where $\tenB^* = \tenC\times_1\matU_1\times_2\matU_2\times_3\matU_3 + \tenE_1$ with $\tenC\in\mathbb{R}^{r\times r\times r}$ being a core tensor with i.i.d. entries from a Gaussian mixture model: $0.5\cdot\mathcal{N}(1,0.5)+0.5\cdot\mathcal{N}(-1,0.5)$, and $\matU_1,\matU_2,\matU_3\in \{0,1\}^{d\times r}$ with only a single $1$ in each row and the noise tensor $\tenE_1\overset{i.i.d.}{\sim}\mathcal{N}(0,0.2)$. We choose this model for generating $\tenB^*$ to ensure that $\tenB^*$ has a checkerboard structure, as shown in Figure~\ref{fig:simulation-setup}, and note that the noiseless part of $\tenB^*$ is also of low tensor-train rank. We re-scale the simulated $\tenB^*$ such that $\maxnorm{\tenB^*}=2$. We enforce each column of $\matU_1,\matU_2,\matU_3$ to have $1$s in consecutive rows so that the simulated $\tenB^*$ demonstrates a noisy ``checker box" structure, as illustrated in Figure~\ref{fig:simulation-setup}(a).

Given the simulated $\tenB^*$, we then simulate the binary data missingness tensor $\tenW$ from the Ising model. Throughout this section, we suppose that two tensor entries $i$ and $j$ are neighbors, i.e., $i\sim j$, if and only if their indices differ by 1 in just one mode. Consequently, for 3-way tensors, each non-boundary entry has six neighbors. We simulate $\tenW$ from the missing propensity model specified by~\eqref{eq:boltzmann-distribution} and~\eqref{eq:hamiltonian} with a block-Gibbs sampler and generate samples from a Monte Carlo Markov Chain (MCMC). The MCMC has $4\times 10^{4}$ iterations with the first $10^{4}$ samples burnt in, and we take one sample every other $10^3$ iterations to end up with $n=30$ samples. In Figure~\ref{fig:simulation-setup}(b), we visualize one simulated $\tenW$.

Lastly, the data tensor $\tenX$ is generated from an additive noise model: $\tenX = \tenX^* + \tenE$, which is similar to $\tenB^*$, with $\tenX^*$ having a Tucker rank $(3,3,3)$. The noiseless tensor $\tenX^*$ also possesses a ``checker box" structure and is contaminated by the noise tensor $\tenE$, whose distribution depends on the specific simulation setting described later. We re-scale $\tenX^*$ to have $\maxnorm{\tenX^*}=2$ and define the signal-to-noise ratio (SNR) of $\tenX$ as $\maxnorm{\tenX^*}/\maxnorm{\tenE}$ and re-scale $\tenE$ such that SNR$=2$. The data tensor $\tenX$ is then masked by $\tenW$, as plotted in Figure~\ref{fig:simulation-setup}(c).

\subsection{Conformal Prediction Validation}\label{subsec:conformal-simulation}
To validate the efficacy of the proposed conformalized tensor completion (CTC) algorithm, we consider the simulation setting with $d\in\{40,60,80,100\}$, $r\in\{3,5,7,9\}$, $g(x,y)\in\{0,xy/15\}$. The noise tensor $\tenE$ is simulated based on two different uncertainty regimes: 1) constant noise: $[\tenE]_s\overset{i.i.d.}{\sim}\mathcal{N}(0,1)$; 2) adversarial noise: $[\tenE]_s$ follows independent Gaussian distribution $\mathcal{N}(0,\sigma_s^2)$, with $\sigma_s = [2\exp(\tenB^*_s)/[1+\exp(\tenB^*_s)]^{-1}$. The adversarial noise simulates cases where the missing entries have higher uncertainty than the observed entries.

\begin{figure}[!htb]
    \centering
    \includegraphics[width=0.98\textwidth]{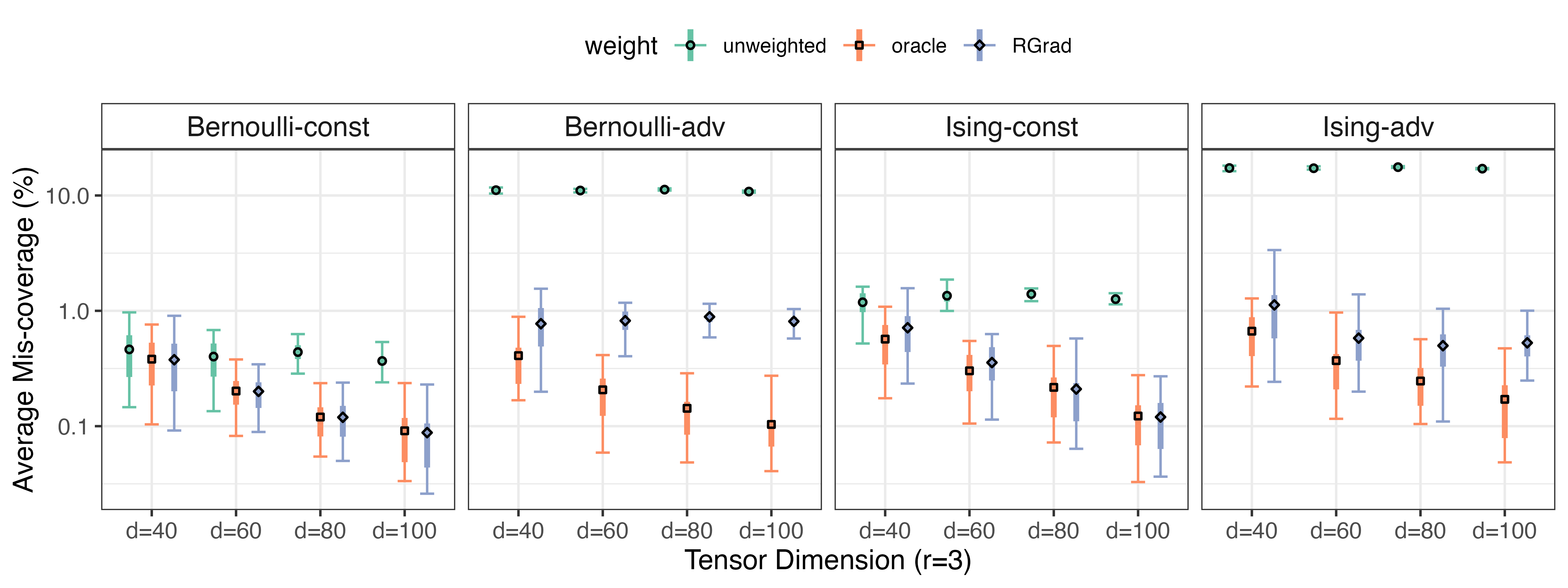}
    \caption{The average mis-coverage of three conformal prediction methods with $d\in\{40,60,80,100\}$, $r=3$ under the Bernoulli and Ising model. Two uncertainty regimes: constant noise (const) and adversarial noise (adv) are considered. Results are based on $30$ repetitions, error bars show the $2.5\%,97.5\%$ quantiles, and the thicker lines show the range of $25\%$ to $75\%$ quantiles. The y-axis is plotted in log10-scale.}
    \label{fig:sim-conformal-coverage}
\end{figure}

For each simulation scenario, we apply the correctly specified CTC algorithm with P-AIC selected rank and call it \textbf{RGrad}. As a benchmark, we also consider two other versions of conformal inference: 1) \textbf{unweighted}: the unweighted conformal prediction; 2) \textbf{oracle}: the weighted conformal prediction with the true tensor parameter $\tenB^*$. We conduct a simulation over $n=30$ repetitions, and for each repetition, we randomly split the observed entries into a training and a calibration set with $q=0.7$ and evaluate the constructed conformal intervals on the missing entries, denoted as $\setS_{miss}$. For the tensor completion algorithm, we choose low Tucker rank tensor completion coupled with Riemannian gradient descent~\citep{wang2023implicit}. We use the absolute residual $\mathcal{S}(y,\widehat{y})=|y-\widehat{y}|$ as the non-conformity score. To evaluate the conformal intervals, we define the average mis-coverage metric as:
\begin{equation}\label{eq:coverage-l1-loss}
    \text{Average Mis-coverage \%} = \frac{100}{|\mathbb{Q}|}\sum_{\tau\in\mathbb{Q}} \left|\tau - \frac{1}{|\setS_{miss}|} \sum_{s\in\setS_{miss}}\indicator{\tenX_s\in\widehat{C}_{\tau,s}(\widehat{\tenX})} \right|,
\end{equation}
with $\mathbb{Q}=\{0.80,0.81,\ldots,0.98,0.99\}$. We plot the average mis-coverage with $r=3$ in Figure~\ref{fig:sim-conformal-coverage}. We also plot the results with $r=9$ in Section~\ref{app:simulation-results} of the supplemental material.

According to the results, we find that with constant entry-wise uncertainty, even the unweighted conformal intervals perform decently, but still have more mis-coverage than the oracle case. Using our CTC algorithm significantly shrinks the mis-coverage and matches the performance of the oracle case. Under the adversarial noise regime, we observed significant mis-coverage ($>10\%$) of the unweighted conformal prediction, and using the CTC algorithm provides conformal intervals with $<1\%$ of mis-coverage, indicating that our method helps in constructing well-calibrated confidence intervals.

The mis-coverage is even worse for the unweighted conformal prediction when missingness is locally dependent based on the Ising model, and the CTC algorithm still provides conformal intervals at the target coverage. In Figure~\ref{fig:sim-coverage-additional-result} of Section~\ref{app:simulation-results} of the supplemental material, we further show that the mis-coverage of the unweighted conformal prediction is mainly under-coverage, as it cannot account for the increase of uncertainty in the testing set under adversarial noise.

\begin{figure}[!htb]
    \centering
    \includegraphics[width=0.98\textwidth]{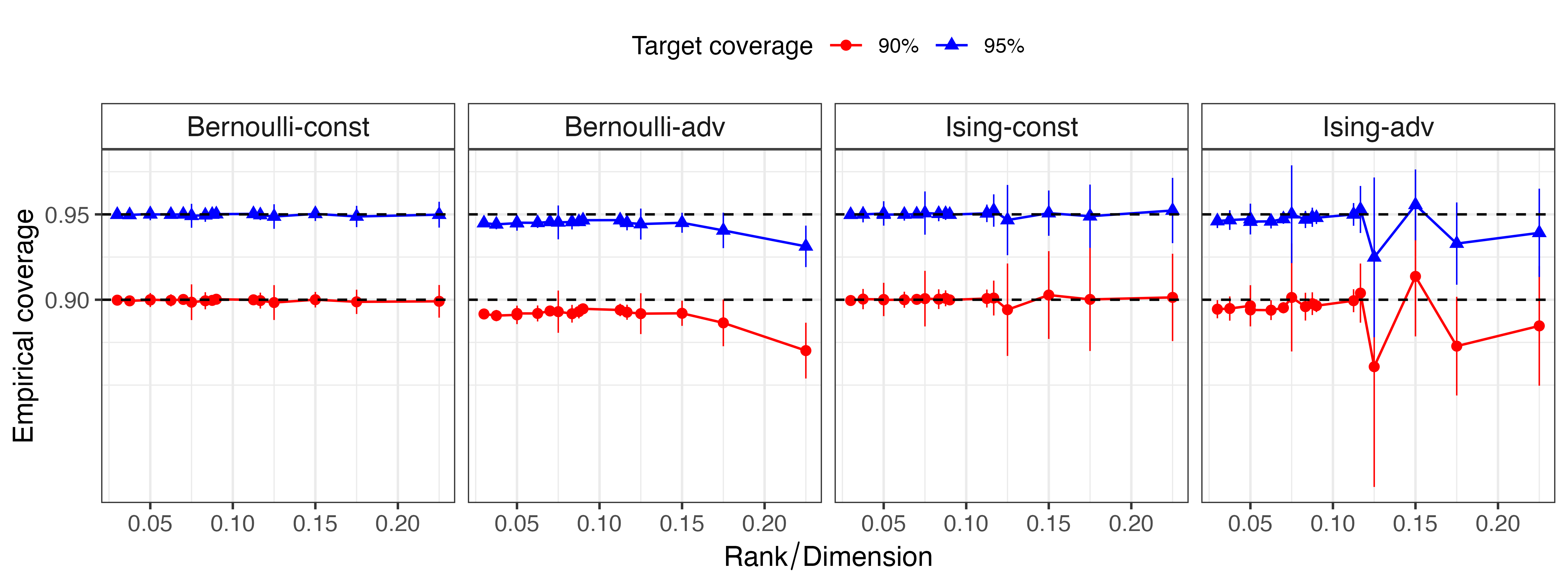}
    \caption{RGrad empirical coverage of the $90\%$ and $95\%$ conformal intervals under the Bernoulli and Ising model with two noise regimes. x-axis is the $r/d$ of the tensor parameter $\tenB^*$. Results are based on $n=30$ repetitions, and error bars are $\pm 1.96$ standard deviations.}
    \label{fig:sim-conformal-coverage-rnd}
\end{figure}

To provide a full landscape on how the conformal intervals based on our CTC algorithm perform under different tensor rank $r$ and tensor dimension $d$ of the underlying parameter $\tenB^*$, we visualize in Figure~\ref{fig:sim-conformal-coverage-rnd} the empirical coverage of $90\%$ and $95\%$ conformal intervals under different missingness and uncertainty regimes by $r/d$, i.e. the rank-over-dimension of the tensor $\tenB^*$, based on our RGrad method. Generally speaking, the higher $r/d$ is, the more difficult it is to estimate the missing propensity of the tensor data and thus the worse the coverage of the conformal intervals, which echoes our theoretical result in Section~\ref{app:cov-guarantee} of the supplemental material. Therefore, we conclude that our proposed method would provide well-calibrated conformal intervals when the underlying missingness model has a low tensor rank relative to the tensor size (i.e., $r << d$).

In Section~\ref{app:simulation-results} of the supplemental material, we also compare our RGrad approach with other binary tensor decomposition approaches, such as CP and Tucker decomposition, for estimating the missing propensity and conducting conformal prediction. We find our method performs consistently well under all kinds of dependency and uncertainty regimes. In Section~\ref{app:ncs-sim} of the supplemental material, we further explore other choices of the non-conformity score, including two-sided and normalized scores, and verify the performance of conformal prediction under these settings.

\section{Data Application to TEC Reconstruction}\label{sec:application}
Our proposed method can account for the locally dependent data missingness, which is a common data missing pattern for spatial data; therefore, we apply our method to a spatio-temporal tensor completion problem in this section as an application. Specifically, we consider the total electron content (TEC) reconstruction problem over the territory of the USA and Canada. The TEC data have severe missing data problems since they can be measured only if the corresponding spatial location has a ground-based receiver. An accurate prediction of the TEC can foretell the impact of space weather on the positioning, navigation, and timing (PNT) service~\citep{wang2021,Younas2022}.  Existing literature~\citep{pan2021tec,sun2022matrix,wang2023forecast} focuses on imputation and prediction of the global and regional TEC and lacks data-driven approaches for quantifying the uncertainty of the imputation, and we aim at filling in this gap.

\begin{figure}[!htb]
    \centering
    \includegraphics[width=\textwidth]{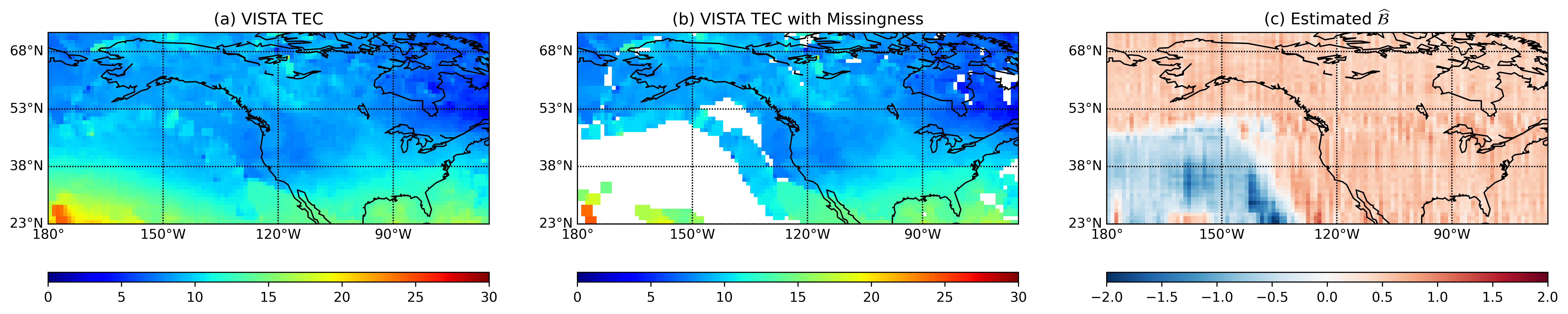}
    \caption{(a) The VISTA TEC at 00:02:30 UT, September 1, 2017. (b) The VISTA TEC in (a) with data missingness from the Madrigal TEC. (c) Fitted $\htenB$ based on the Ising model.}
    \label{fig:TEC-plot}
\end{figure}

In Figure~\ref{fig:TEC-plot}(a), we plot the TEC distribution over the USA and Canada from the VISTA TEC database~\citep{sun2023complete}. The VISTA TEC is a pre-imputed version of the Madrigal TEC~\citep{MadrigalTEC}, which has $>80\%$ of the data missing globally. We use the VISTA TEC as the ground truth and the Madrigal TEC data missingness to mask out entries in the VISTA TEC to simulate data missingness close to what scientists observe in practice. 

To set up the experiment, we use the first $20$ days of data in September 2017, and each day consists of a tensor of size $50\times 115 \times 96$. We use the first $5$ days as a validation set to search for the best $g(\cdot,\cdot)$ function for the Ising model. For each day, we fit the CTC algorithm with a simple tensor completion algorithm based on~\eqref{eq:tensor-completion-problem} with a Tucker rank at $(3,3,3)$ and pick the tensor-train rank $\vecr = (r,r)$ by P-AIC. Based on Figure~\ref{fig:sim-conformal-coverage-rnd}, we know that the Ising model exhibits under-coverage as $r/d$ increases over $0.15$; therefore, we select the rank $r$ from $2\le r\le 7$ only. For each day, we consider the Ising model with $g(x,y)=5xy/4, h(x)=x/2$, the Bernoulli model with $g(x,y)=0,h(x)=x/2$, and the unweighted conformal prediction for comparison. In Table~\ref{tab:application-numerical-results}, we report the results on the average mis-coverage \% and the empirical coverage of $90\%$ and $95\%$ CI.
\begin{table}[!htb]
    \centering
    \begin{tabular}{|c|c|c|c|}
    \hline
    method & mis-coverage \% & 90\% CI coverage \% & 95\% CI coverage \% \\\hline
    unweighted & $42.1(6.49)$ & $46.3(6.58)$ & $52.3(7.23)$ \\
    Bernoulli & $23.1(5.34)$ & $64.6(5.97)$ & $76.8(5.03)$ \\
    Ising & $6.01(2.45)$ & $90.0(6.06)$ & $94.2(3.74)$\\\hline
    \end{tabular}
    \caption{Mis-coverage \% and empirical coverage of CI at $90\%$ and $95\%$ level for the unweighted conformal prediction and weighted conformal prediction with Bernoulli and Ising model for data during Sept 6 to Sept 20, 2017.}
    \label{tab:application-numerical-results}
\end{table}

\begin{figure}[!htb]
    \centering
    \includegraphics[width=\textwidth]{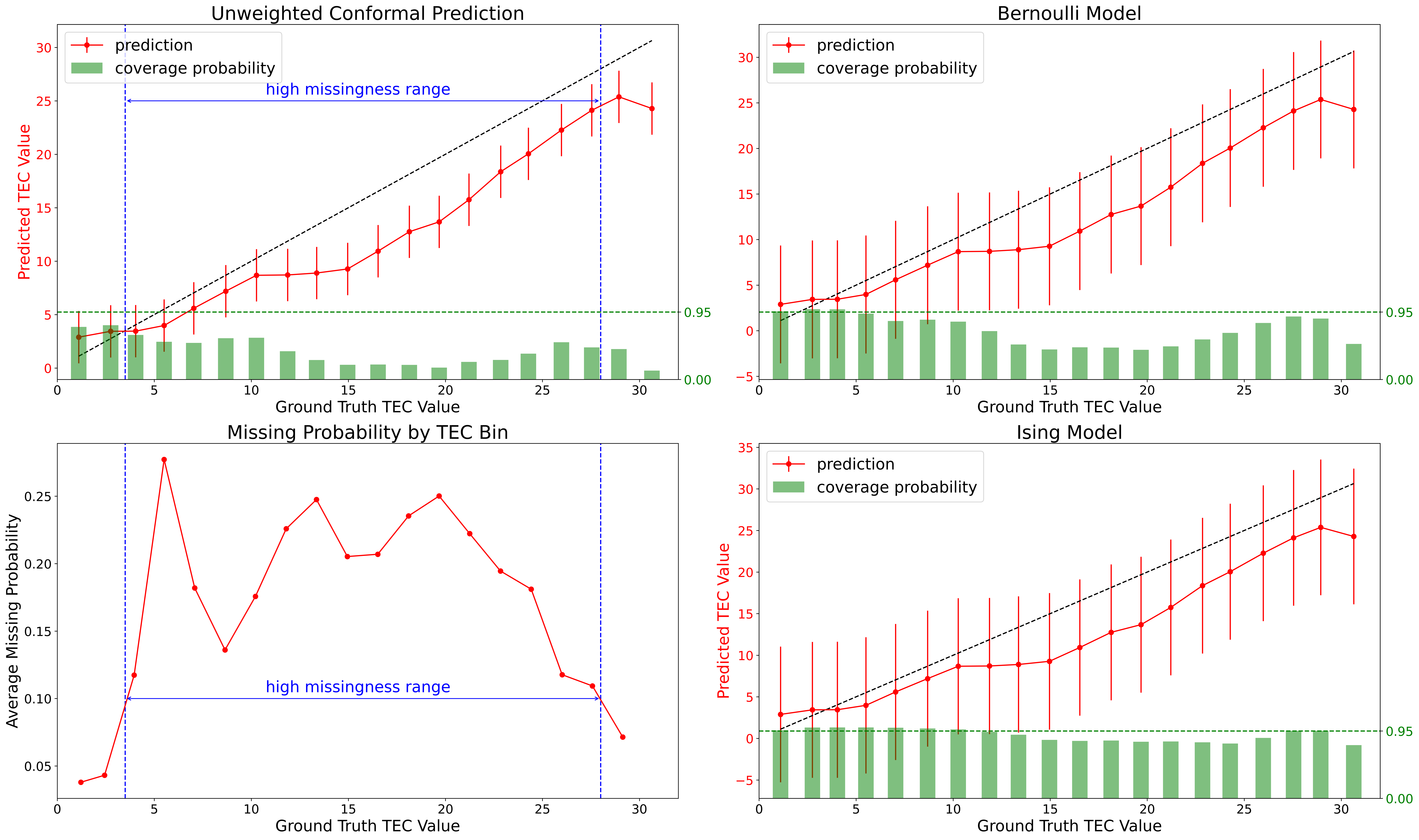}
    \caption{All except the lower-left panels show the average $95\%$ conformal intervals and the empirical coverage for 20 different bins of TEC values on Sept 6, 2017. Each bin spans 1.5 TEC units. The lower-left panel shows the missing probability of different bins. A bin is termed ``high missingness" if $>10\%$ of the data is missing.}
    \label{fig:application-TEC-CI}
\end{figure}

In Figure~\ref{fig:application-TEC-CI}, we visualize the average $95\%$ CI and its empirical coverage for $20$ different bins of TEC values on Sept 6, 2017. It is shown that the data missingness is not uniform across different bins of TEC values, and different bins have different distributions of the imputation errors (see how the prediction deviates from the truth), making the unweighted conformal prediction less favorable, especially when data missingness is high. These empirical results reveal that by accounting for the heterogeneity and the spatial dependency of data missingness, one can construct well-calibrated confidence intervals using our method.

\section{Conclusion}\label{sec:conclusion}
In this paper, we propose a data-driven approach for quantifying the uncertainty of tensor completion. Our method consists of two major steps. We first estimate the missing propensity of each tensor entry using a parameterized Ising model with a low tensor-train rank parameter, and then plug in the missing propensity estimator to weight each tensor entry, and then construct the confidence region with split conformal prediction. We implement the estimation of missing propensity with a computationally efficient Riemannian gradient descent algorithm and validate the resulting conformal intervals with extensive simulation studies and an application to regional TEC reconstruction. We focus on tensor-train rank in our implementation, but our method can be easily extended to other tensor ranks as long as the low-rank tensors lie on a smooth manifold.

There are two limitations of our method. Firstly, we do not have a systematic approach to determine the best specification of the Ising model, e.g., the hyperparameters for the function $g(x,y)$ and $h(x)$. We recommend using a joint gradient descent approach, where one uses RGrad for $\tenB$ and gradient descent for the hyperparameters. Alternatively, one can also try grid search on a held-out validation set.

Secondly, our Ising model can only account for locally dependent missingness, but not arbitrary missingness. In our simulation experiment, we consider adversarial noise, which is somewhat similar to the missing not-at-random (MNAR) scenario where the uncertainty of the entry is related to the missing propensity. Our model can be extended to handle more flexible missingness, such as those in~\citet{tabouy2020variational,sportisse2020imputation}. We leave these topics to future research.

\section*{Acknowledgement}
We thank Shasha Zou for helpful discussions on the TEC data. YC is supported by NSF DMS 2113397, NSF PHY 2027555, NASA Federal Award No. 80NSSC23M0192 and No. 80NSSC23M0191. 

\bibliographystyle{agsm}
\bibliography{references}

\newpage
\appendix
\renewcommand{\theequation}{S.\arabic{equation}}
\renewcommand{\thefigure}{S.\arabic{figure}}
\renewcommand{\thetable}{S.\arabic{table}}
\setcounter{equation}{0}
\setcounter{figure}{0}
\setcounter{table}{0}

\begin{center}
{\large\bf SUPPLEMENTARY MATERIAL}
\end{center}
This supplemental material contains three sections. Section~\ref{app:proof-main} contains the proofs for proposition~\ref{thm:conformal-interval-weight} and additional theoretical results for the estimation error and coverage guarantee for the Bernoulli model. Section~\ref{app:lemma} describes the technical lemmas used in Section~\ref{app:proof-main}. Section~\ref{app:simulation} contains additional details and results of the simulation experiments. All figures, tables, and equations in the supplemental material are numbered with a prefix ``S", which distinguishes them from the main paper numbering.

\addcontentsline{toc}{section}{Appendix}
\startcontents[appendix]
\printcontents[appendix]{}{1}{\section*{Contents}}

\section{Proofs of Theorems and Propositions}\label{app:proof-main}
Throughout this section, for any tensor $\tenB\in\mathbb{R}^{\dk}$, we use $\bar{d}, d^*$ to denote $\sum_k d_k$ and $\prod_k d_k$, respectively. For any tensor-train rank $\vecr=(r_1,\ldots,r_{K-1})$, we use $r^*$ to denote $\prod_k r_k$. We use $c, c^{\prime}, C, C_0, C_1,\ldots$ to denote positive absolute constants and $c_{K}, c^{\prime}_K, C_K, C_{K,0}, C_{K,1},\ldots$ to denote positive constants that only relate to $K$. For two sequences $\{a_n\}_{n=1}^{\infty}$ and $\{b_n\}_{n=1}^{\infty}$, we use $a_n\asymp b_n$ to represent $\lim_{n\rightarrow\infty} a_n/b_n = C > 0$, with $C$ being finite.

\subsection{Proof of Proposition~\ref{thm:conformal-interval-weight}}\label{app:proof-weighted-conformal-coverage}
\begin{proof}
Given any testing entry $s^*\in\setS_{miss}$, we relabel all elements in $\setS_{cal} \cup \{s^*\}$ as $\{s_1,\ldots,s_{n+1}\}$. Now recall the definition of $\tenE_0$ as:
\begin{equation*}
    \tenE_0=\left\{\ttenW_s=1 \text{ for } s\in\setS_{tr}\cup\setS_{cal},  \setS_{cal}\cup\{s^*\}=\{s_1,\ldots,s_{n+1}\} \text{ and }\ttenW_s=-1 \text{ o.w.}\right\},
\end{equation*}
namely one observes data only at $\setS_{tr}$ and $n$ out of $n+1$ entries from $\{s_1,\ldots,s_{n+1}\}$. 

Let $V$ denote the non-conformity score of the testing entry, then the weighted exchangeability framework in~\citet{tibshirani2019conformal} states that one can treat $V$ as a weighted draw from $\{\mathcal{S}(\tenX_{s_1},\htenX_{s_1}),\ldots,\mathcal{S}(\tenX_{s_{n+1}},\htenX_{s_{n+1}})\}$, with weight being:
\begin{equation*}
    \Prob\left[V=\mathcal{S}(\tenX_{s_k},\htenX_{s_k})\bigg|\tenE_0\right] = \frac{\Prob\left[\ttenW_{s_k}=-1, \ttenW_s=1 \text{ for }s\in\setS_k \bigg|\tenE_0\right]}{\sum\limits_{l=1}^{n+1}\Prob\left[\ttenW_{s_l}=-1, \ttenW_s=1 \text{ for }s\in\setS_l  \bigg|\tenE_0\right]},
\end{equation*}
where $\setS_k=\{s_1,\ldots,s_{n+1}\}\setminus\{s_k\}$, for $k=1,\ldots,n+1$. Multiplying both the numerator and the denominator by $\Prob(\tenE_0)$ leads to the weight in the form of $p_k / \sum_{l=1}^{n+1}p_l$, with $p_k$ defined as~\eqref{eq:pk-definition}. The coverage guarantee in~\eqref{eq:weighted-conformal-coverage-guarantee} is then a direct result of Theorem 2 of~\citet{tibshirani2019conformal}.
\end{proof}

\subsection{Approximation Error of Conformal Weights}\label{app:weight-error-bound}
In this subsection, we estimate the additional errors introduced by the approximation we made when computing the conformal weight in~\eqref{eq:Ising-model-conformal-weight-exact}. Basically, to eliminate the dependency of $\omega_k(s^*)$ on the specific $s^*\in\setS_{miss}$, we modify~\eqref{eq:Ising-model-conformal-weight-exact} as:
\begin{equation}\label{eq:conformal-weight-approx}
    \omega_k = \frac{p_k}{\sum_{i=1}^{n+1} p_i} = \frac{\exp\left[-2\sum_{s_j\in\mathcal{N}(s_k)} g(\tenB_{s_k},\tenB_{s_j})\tenW_{s_j}-2h(\tenB_{s_k})\right]}{\sum_{i=1}^{n+1} \exp\left[-2\sum_{s_j\in\mathcal{N}(s_i)} g(\tenB_{s_i},\tenB_{s_j})\tenW_{s_j}-2h(\tenB_{s_i})\right]},
\end{equation}
i.e. setting $\ttenW_{s_{n+1}} = -1$ instead of $1$. We summarize the result in the proposition below.

\begin{proposition}\label{thm:approx-weight-error}
Let $s_1,\ldots,s_n\in \setS_{cal}$, $s_{n+1}=s^*\in\setS_{miss}$, and let $F^*(\cdot)$ be the CDF of the distribution:
$$
\sum_{i=1}^n \omega_{i}(s^*)\cdot\delta_{\mathcal{S}(\tenX_{s_i},\htenX_{s_i})} + \omega_{n+1}(s^*)\cdot\delta_{+\infty},
$$
with $\omega_i(s^*)$ defined in~\eqref{eq:Ising-model-conformal-weight-exact}.
Similarly, let $F(\cdot)$ be the CDF of the distribution:
$$
\sum_{i=1}^n \omega_{i}\cdot\delta_{\mathcal{S}(\tenX_{s_i},\htenX_{s_i})} + \omega_{n+1}\cdot\delta_{+\infty},
$$
with $\omega_i$ defined in~\eqref{eq:conformal-weight-approx}. Let $\setN_{0}=\{i\in[n]|s_i\in\mathcal{N}(s_{n+1})\}$, and $\gamma(\tenB) = \min_{j\in\setN_0} g(\tenB_{s_j}, \tenB_{s_{n+1}})$. Then we have the following universal bound over $|F^*(x) - F(x)|$:
\begin{equation}\label{eq:CDF-deviation-bound}
\sup_{x\in\mathbb{R}} \left|F^*(x) - F(x)\right| \le 3\cdot\max\left\{1, e^{4\gamma(\tenB)}-1\right\}\cdot\sum_{k\in\setN_0} \omega_{k}(s^*).
\end{equation}
\end{proposition}

\begin{proof}
Define $\ta_k$ as $\exp\left[-2\sum_{s_j\in\mathcal{N}(s_k)} g(\tenB_{s_k},\tenB_{s_j})\ttenW_{s_j}-2h(\tenB_{s_k})\right]$, and define $a_k$ similar to $\ta_k$ but replace $\ttenW_{s_j}$ with $\tenW_{s_j}$. Then we have $\omega_k = a_k/\sum_i a_i$ and $\omega_k(s^*)=\ta_k/\sum_i \ta_i$. By definition, $\ta_i = a_i$ if and only if $i\notin \setN_0$. Then we have for any $l=1,\ldots,n+1$:
\begin{align}
\left|\omega_l(s^*) - \omega_l\right| & = \left|\frac{\ta_l}{\sum_{j\in \setN_0} \ta_j + \sum_{j\notin \setN_0}a_j} - \frac{a_l}{\sum_{j\in \setN_0} a_j + \sum_{j\notin \setN_0}a_j}\right|\nonumber \\
& = \left|\frac{(\ta_l-a_l)\cdot\sum_{j\notin \setN_0}a_j + (\ta_l-a_l)\cdot\sum_{j\in \setN_0}a_j + a_l\cdot\sum_{j\in\setN_0}\left(a_j-\ta_j\right)}{\left(\sum_{j\in \setN_0} \ta_j + \sum_{j\notin \setN_0}a_j\right)\left(\sum_{j\in \setN_0} a_j + \sum_{j\notin \setN_0}a_j\right)}\right| \nonumber \\
& \le 2\cdot \frac{|\ta_l - a_l|}{\sum_{j\in \setN_0} \ta_j + \sum_{j\notin \setN_0}a_j} + \omega_l\cdot\sum_{j\in\setN_0} \frac{|\ta_j - a_j|}{\sum_{j\in \setN_0} \ta_j + \sum_{j\notin \setN_0}a_j} \nonumber \\
& = 2\cdot\omega_l(s^*)\cdot\left(1 - \exp\left[-4\cdot\indicator{l\in\setN_0}\cdot g(\tenB_{s_l},\tenB_{s_{n+1}})\right]\right) \nonumber \\
& + \omega_l\cdot \sum_{j\in\setN_0} \omega_j(s^*)\cdot \left(1 - \exp\left[-4\cdot g(\tenB_{s_j},\tenB_{s_{n+1}})\right]\right) \nonumber \\
& \le \max\left\{1, e^{4\gamma(\tenB)-1}\right\}\cdot \left[2\cdot\omega_l(s^*)\cdot \indicator{l\in\setN_0} + \omega_l\cdot \sum_{j\in\setN_0} w_{j}(s^*)\right].
\end{align}

Then for any $x\in\mathbb{R}$, we have:
\begin{align}
\left|F^*(x) - F(x)\right| & = \left|\sum_{i:\mathcal{S}(\tenX_{s_i}, \htenX_{s_i})\le x} (\omega_i(s^*) - \omega_i)\right| \nonumber \\ 
& \le \sum_{i:\mathcal{S}(\tenX_{s_i}, \htenX_{s_i})\le x} \left|\omega_i(s^*) - \omega_i\right| \nonumber \\ 
& \le \sum_{i:\mathcal{S}(\tenX_{s_i}, \htenX_{s_i})\le x} \max\left\{1, e^{4\gamma(\tenB)-1}\right\}\cdot \left[2\cdot\omega_i(s^*)\cdot \indicator{i\in\setN_0} + \omega_i\cdot \sum_{j\in\setN_0} w_{j}(s^*)\right] \nonumber \\ 
& \le \max\left\{1, e^{4\gamma(\tenB)-1}\right\} \cdot \left[2\cdot\sum_{i\in\setN_0} \omega_i(s^*) + \left(\sum_{i:\mathcal{S}(\tenX_{s_i}, \htenX_{s_i})\le x} \omega_i\right)\cdot \sum_{j\in\setN_0} \omega_j(s^*)\right] \nonumber \\
& \le 3\cdot\max\left\{1, e^{4\gamma(\tenB)}-1\right\}\cdot\sum_{k\in\setN_0} \omega_{k}(s^*). \nonumber
\end{align}
\end{proof}

Proposition~\ref{thm:approx-weight-error} provides a universal upper bound over the empirical CDF of the distribution of non-conformity scores under the exact and approximated weights. The upper bound showed that the deviation is determined by the sum of the weights at all entries that are neighbors of $s^*$. Typically, we specify the neighbors of each tensor entry as all entries whose indices differ on only one dimension. So for an order-3 tensor, each entry has at most $6$ neighbors, but the size of the calibration set is much larger, leading to generally a very small deviation of the empirical CDFs. The bound in~\eqref{eq:CDF-deviation-bound} can be further refined given specific $\tenB$, but we leave this general conclusion here to show the minimal impact of the approximation step.

\subsection{Bernoulli Model Estimation Error Bound}\label{app:error-bound}
In this subsection, we derive the error bound of the MPLE estimator $\htenB$ under the assumption that $g(x,y)=0$, i.e., all entries of $\trtenW$ are observed independently with probability $q(\exp[-2h(\tenB^*_s)]+1)^{-1}$. Evidently, under this assumption, the MPLE is identical to MLE since the pseudo-likelihood is also the true Bernoulli likelihood. Our main result is in Theorem~\ref{thm:error-bound-Bernoulli}. To establish the theoretical result, we make several additional assumptions:
\begin{assumption}\label{assp:h-function-convex}
$h(\cdot): \mathbb{R}\mapsto\mathbb{R}$ is a non-decreasing, non-constant twice continuously differentiable function with $h^{\prime\prime}(\cdot) \ge 0$.
\end{assumption}
\begin{assumption}\label{assp:B-boundness}
The MPLE estimator $\htenB$ and the true tensor parameter $\tenB^*$ have bounded max-norm: $\maxnorm{\htenB}, \maxnorm{\tenB^*} \le \xi$.
\end{assumption}
We define $f(x) = \exp[2h(x)]/(1 + \exp[2h(x)])$ and the following two  constants:
\begin{equation*}
    \alpha_{\xi} = \sup_{|x|\le\xi} |2h^{\prime}(x)|, \quad \gamma_\xi = \inf_{|x|\le\xi} \min\left\{\left[\frac{f^{\prime}(x)}{f(x)}\right]^2-\frac{f^{\prime\prime}(x)}{f(x)},\frac{qf^{\prime\prime}(x)}{1-qf(x)}+\left[\frac{qf^{\prime}(x)}{1-qf(x)}\right]^2\right\}.
\end{equation*}
To see what these two constants represent, recall that the negative log-likelihood for $\trtenW$ given $\tenB$ can be written as the sum of each entry's negative log-likelihood $\ell_i([\trtenW]_i|\tenB)$, which is defined as:
\begin{equation*}
    \ell_i([\trtenW]_i|\tenB) = -\left[\left(\frac{[\trtenW]_i+1}{2}\right)\log qf(\tenB_i) + \left(\frac{1-[\trtenW]_i}{2}\right)\log (1-qf(\tenB_i))\right].
\end{equation*}
It is not difficult to verify that $\alpha_\xi$ upper bounds $|\partial \ell_i(\cdot|\tenB)/\partial \tenB_i|$ and $\gamma_\xi$ lower bounds $\partial^2 \ell_i(\cdot|\tenB)/\partial\tenB_i^2$ for all $i$ as long as $\max_s |\tenB_s|\le\xi$. By excluding the trivial case where $h(\cdot)$ is a constant function, $\alpha_\xi$ is strictly positive. If for all $|x|\le\xi$, we have $1-(1-q)f(x)-f^2(x) > 0$, then we can verify that $\gamma_\xi > 0$ for common choices of $h(\cdot)$, such as the logit model $h(x)=x/2$ or the probit model $h(x)=2^{-1}\log[\Phi(x)/(1-\Phi(x))]$. For the remainder of the appendix, we will assume generally that $\gamma_\xi > 0$, which is simply saying that the function $\ell_i(\cdot|\tenB)$ is $\gamma_\xi$-strongly convex.

Finally, it is useful to define the tensor spectral norm and the tensor nuclear norm here:
\begin{definition}\label{def:tensor-spec-norm}
For a tensor $\tenA\in\mathbb{R}^{\dk}$, its spectral norm, denoted as $\|\tenA\|_\sigma$, is defined as:
\begin{equation*}
\|\tenA\|_\sigma = \sup_{\vecu_1,\ldots,\vecu_K} \left\langle\tenA,\vecu_1 \circ\cdots\circ\vecu_K\right\rangle, \quad \vecu_k\in\setS^{d_k-1}, \forall k,    
\end{equation*}
where $\circ$ denotes vector outer product and $\setS^{d_k-1}$ is a unit sphere in $\mathbb{R}^{d_k}$.
\end{definition}
\begin{definition}\label{def:tensor-nuclear-norm}
For a tensor $\tenC\in\mathbb{R}^{\dk}$, its nuclear norm $\|\tenC\|_*$ is defined as:
\begin{equation*}
    \|\tenC\|_* = \inf\left\{\sum_{r}\lambda_r\bigg| \tenC = \sum_r \lambda_r \vecu_1\circ\cdots\circ\vecu_K, \vecu_k\in\setS^{d_k-1}, \forall k\right\}.
\end{equation*}
\end{definition}

With the aforementioned assumptions and notations, we have the following non-asymptotic bound on $\twonorm{\htenB-\tenB^*}$:
\begin{theorem}\label{thm:error-bound-Bernoulli}
Assume that $g(x,y)=0$ and assumption~\ref{assp:h-function-convex} and~\ref{assp:B-boundness} hold, and further assume that $\htenB$ reaches the global minimum of the negative log-likelihood $\ell(\trtenW|\tenB)$ and the entry-wise negative log-likelihood is $\gamma_\xi$-strongly convex with $\gamma_\xi > 0$, then:
\begin{equation}\label{eq:estimator-error-bound}
    \Prob\left(\frac{1}{\sqrt{d^*}}\twonorm{\htenB-\tenB^*} \le 2C_{K,1}\frac{\alpha_\xi}{\gamma_{\xi}}\sqrt{\frac{r^*\bar{d}}{d^*}}\right) \ge 1-\exp\left(-C_1\bar{d}\log K\right),
\end{equation}
where $C_1,C_{K,1}$ are some positive constants.
\end{theorem}
\begin{proof}
Using Taylor expansion upon $\ell(\trtenW|\htenB)$ at $\tenB=\tenB^*$ yields:
\begin{equation}\label{eq:second-order-Taylor-expansion}
    \ell(\trtenW|\htenB) = \ell(\trtenW|\tenB^*) + \left\langle\nabla\ell(\trtenW|\tenB^*), \htenB-\tenB^*\right\rangle + \frac12 \vect{\htenB-\tenB^*}^{\top}\mathbf{H}(\check{\tenB})\vect{\htenB-\tenB^*},
\end{equation}
where $\check{\tenB}$ is a convex combination of $\htenB$ and $\tenB^*$. Since, by assumption, $\htenB$ reaches the global minimum of $\ell(\trtenW|\tenB)$, or $\ell(\tenB)$ in short, we have $\ell(\htenB) \le \ell(\tenB^*)$, and thus the sum of the last two terms in~\eqref{eq:second-order-Taylor-expansion} are no greater than zero.

For the first term, let $\tenG^* = \nabla\ell(\tenB^*)$ and $\tenG^*$ satisfies:
\begin{equation}
    [\tenG^*]_s = -[1-f(\tenB^{*}_s)]\cdot 2h^{\prime}(x)\cdot\indicator{[\trtenW]_s=1} + \frac{qf(\tenB^{*}_s)[1-f(\tenB^{*}_s)]}{1-qf(\tenB^{*}_s)}\cdot 2h^{\prime}(x)\cdot\indicator{[\trtenW]_s=-1},
\end{equation}
and it is easy to verify that $\mathrm{E}[[\tenG^*]_s]=0$ and $\maxnorm{\tenG^*} \le \alpha_{\xi}$. By Lemma~\ref{lemma:inner-product-bound}, we can lower bound the first term as:
\begin{equation}\label{eq:Taylor-expansion-lb-1}
    \left\langle\nabla\ell(\trtenW|\tenB^*), \htenB-\tenB^*\right\rangle \ge -\|\tenG^*\|_\sigma\|\htenB-\tenB^*\|_*.
\end{equation}
By Lemma~\ref{lemma:rank-bound-tensor-train}, we have $\ranktt{\htenB-\tenB^*}\le 2\vecr$, and then by Lemma~\ref{lemma:nuclear-frobenius-bound}, we have $\|\htenB-\tenB^*\|_* \le \sqrt{(2r_1)\cdots(2r_{K-1})}\cdot\twonorm{\htenB-\tenB^*}$. Therefore, to lower bound the RHS of~\eqref{eq:Taylor-expansion-lb-1}, we only need to upper bound the spectral norm of $\tenG^*$. Since entry-wisely, $\tenG^*$ is mean-zero and bounded by $\alpha_\xi$ (therefore the sub-Gaussian norm is $\alpha_\xi$), we can apply Lemma~\ref{lemma:concentration-tensor-spec-norm} and get:
\begin{equation}\label{eq:G-spectral-norm-tail-bound}
    \Prob\left(\|\tenG^*\|_{\sigma} \le \sqrt{8\alpha_{\xi}^2\left[\bar{d}\log 5K+\log\frac{2}{\delta}\right]}\right) \ge 1-\delta.
\end{equation}
By setting $\delta=\exp\left(-C_1\bar{d}\log K\right)$, with $C_1$ be some absolute constant, we can simplify~\eqref{eq:G-spectral-norm-tail-bound} as:
\begin{equation}\label{eq:G-spectral-norm-tail-bound-2}
    \Prob\left(\|\tenG^*\|_{\sigma} \le C_K\alpha_\xi\sqrt{\bar{d}}\right) \ge 1-\exp\left(-C_1\bar{d}\log K\right),
\end{equation}
with $C_K = \sqrt{8\left(\log 5K+C_1\log K + 1\right)}$.

Combining these results, we can lower bound the RHS of~\eqref{eq:Taylor-expansion-lb-1} by:
\begin{equation}\label{eq:Taylor-expansion-lb-2}
-\|\tenG^*\|_\sigma\|\htenB-\tenB^*\|_* \ge  -C_{K,1}\alpha_{\xi}\sqrt{\bar{d}r^*} \twonorm{\htenB-\tenB^*},
\end{equation}
with probability at least $1-\exp\left(-C_1\bar{d}\log K\right)$, where $C_{K,1} = 2^{(K-1)/2}C_K$.

For the quadratic form in~\eqref{eq:second-order-Taylor-expansion}, we have:
\begin{equation}\label{eq:Taylor-expansion-lb-3}
    \frac12 \vect{\htenB-\tenB^*}^{\top}\mathbf{H}(\check{\tenB})\vect{\htenB-\tenB^*} \ge \frac{\gamma_\xi}{2}\twonorm{\htenB-\tenB^*}^2 > 0.
\end{equation}
Combining~\eqref{eq:Taylor-expansion-lb-2} and~\eqref{eq:Taylor-expansion-lb-3}, we obtain:
\begin{equation*}
    \Prob\left(\frac{1}{\sqrt{d^*}}\twonorm{\htenB-\tenB^*} \le 2C_{K,1}\frac{\alpha_\xi}{\gamma_{\xi}}\sqrt{\frac{r^*\bar{d}}{d^*}}\right) \ge 1-\exp\left(-C_1\bar{d}\log K\right),
\end{equation*}
which completes the proof.
\end{proof}

\begin{remark}\label{rmk:estimator-error-bound-remark}
Under the scenario where $d_1 \asymp \cdots \asymp d_K \asymp O(d)$ and $r_1 \asymp \cdots r_{K-1} \asymp O(r)$, the result in~\eqref{eq:estimator-error-bound} can be reduced to:
\begin{equation*}
    \Prob\left(\frac{1}{\sqrt{d^*}}\twonorm{\htenB-\tenB^*} \le 2C_{K}\frac{\alpha_\xi}{\gamma_{\xi}}\sqrt{\left(\frac{r}{d}\right)^{K-1}}\right) \ge 1-\exp\left(-C_1\bar{d}\log K\right).
\end{equation*}
So the estimating error can scale with $r/d$, where a lower $r/d$ generally poses an easier binary tensor decomposition problem with lower root mean-squared error.
\end{remark}

Deriving an error bound similar to Theorem~\ref{thm:error-bound-Bernoulli} is quite infeasible for the Ising model. There are two major theoretical challenges:
\begin{enumerate}
    \item showing the concentration inequality of the spectral norm of the gradient tensor $\tenG$ at the true parameter $\tenB^*$. It is easier to derive it under the assumption that all entries are independent (Lemma~\ref{lemma:concentration-tensor-spec-norm}), but much harder to do so under local dependency.
    \item proving the strong convexity of the negative pseudo-likelihood near the true value $\tenB^*$. Again, when the missingness is completely independent, this is feasible, as we have proved in~\eqref{eq:Taylor-expansion-lb-3}. However, there is no such guarantee under the locally dependent missingness since each entry's missingness depends on the parameters within its neighborhood.
\end{enumerate}
Proving these facts requires a significant amount of extra research, and we leave this for future theoretical works.

\subsection{Bernoulli Model Conformal Inference Coverage Guarantee}\label{app:cov-guarantee}
In this subsection, we utilize the theoretical result in Theorem~\ref{thm:error-bound-Bernoulli} and derive the coverage probability lower bound of the CTC algorithm under the Bernoulli model. The result will reveal how the estimating error of $\tenB^*$ propagates into the mis-coverage rate. To begin with, we state an essential lemma, which is a trivial extension of Theorem 3.2 of~\citet{gui2023conformalized} under the conformalized matrix completion context:
\begin{lemma}[Theorem 3.2 of~\citet{gui2023conformalized}]\label{lemma:coverage-lower-bound}
Let $\htenX$ be the output of any tensor completion algorithm, and $\htenB$ be the output of the RGrad algorithm and both $\htenX,\htenB$ are based on $\setS_{tr}$ only, then given that $g(x,y)=0$, we have:
\begin{equation}\label{eq:CTC-coverage-lb-lemma}
    \mathrm{E}\left[\frac{1}{|\setS_{miss}|}\sum_{s\in\setS_{miss}}\indicator{\tenX_s\in\widehat{C}_{1-\alpha,s}(\htenX)}\right] \ge 1-\alpha - \mathrm{E}[\Delta],
\end{equation}
where $\widehat{C}_{1-\alpha,s}(\htenX)$ is the conformal interval for testing entry $s$ at $(1-\alpha)$ level by the CTC algorithm and $\Delta$ is defined as:
\begin{equation}\label{eq:Delta-definition}
    \Delta = \frac12 \sum_{s\in\setS_{cal}\cup \{s^*\}} \left|\frac{\exp[-2h(\htenB_s)]}{\sum_{s\in\setS_{cal}\cup \{s^*\}}\exp[-2h(\htenB_s)]} - \frac{\exp[-2h(\tenB^*_s)]}{\sum_{s\in\setS_{cal}\cup \{s^*\}}\exp[-2h(\tenB^*_s)]}\right|.
\end{equation}
\end{lemma}
We neglect the proof here since the generalization from matrix to tensor setting is trivial, as one can matricize the tensor into a matrix, and the result holds automatically. By Lemma A.1 in~\citet{gui2023conformalized}, one can further upper bound $\Delta$ by:
\begin{equation}\label{eq:Delta-upper-bound}
    \Delta \le \frac{\|\exp[-2h(\htenB)] - \exp[-2h(\tenB^*)]\|_1}{\sum_{s\in\setS_{cal}}\exp[-2h(\htenB_s)]},
\end{equation}
where the $h(\cdot)$ is applied to tensors element-wisely and $\|\cdot\|_1$ is the element-wise tensor $\ell_1$ norm. The quantity $\Delta$ is trivially bounded by $1$ as it is the total-variation (TV) distance between two CDFs of discrete random variables. With this lemma, we now formally state our main result:
\begin{theorem}\label{thm:CTC-converage-guarantee}
Assume that the same assumptions hold as Theorem~\ref{thm:error-bound-Bernoulli} and further denote $l_{\xi}=\inf_{|x|\le \xi}\exp[-2h(x)], u_{\xi}=\sup_{|x|\le \xi}\exp[-2h(x)]$. The $(1-\alpha)$-level conformal interval $\widehat{C}_{1-\alpha,s}(\htenX)$ satisfies:
\begin{align}
    \mathrm{E}\left[\frac{1}{|\setS_{miss}|}\sum_{s\in\setS_{miss}}\indicator{\tenX_s\in\widehat{C}_{1-\alpha,s}(\htenX)}\right] & \ge 1-\alpha -\frac{2C_{K,1}c_\xi}{(1-c)(1-q)}\sqrt{\frac{r^*\bar{d}}{d^*}} \nonumber\\
    & -\exp[-C_1\bar{d}\log K]- \exp\left[-\frac{c^2(1-q)d^*l_\xi}{2}\right]\label{eq:under-coverage-bound},
\end{align}
for any $0<c<1$, where $q$ is the train-calibration split probability in the CTC algorithm and $c_\xi=u_\xi\alpha_{\xi}^2/(\gamma_\xi l_\xi^2)$.
\end{theorem}
\begin{proof}
Given Lemma~\ref{lemma:coverage-lower-bound}, the coverage guarantee can be derived if one can characterize an upper bound for $\mathrm{E}[\Delta]$. To upper bound $\Delta$, we start from~\eqref{eq:Delta-upper-bound} and bound the numerator on the RHS of~\eqref{eq:Delta-upper-bound} as:
\begin{align}
    \|\exp[-2h(\htenB)] - \exp[-2h(\tenB^*)]\|_1 & \le \sup_{|x|\le\xi} \left|\exp[-2h(x)]\cdot 2h^{\prime}(x)\right| \cdot \|\htenB - \tenB^*\|_1 \nonumber  \\
    & \le u_{\xi}\alpha_\xi\cdot \|\htenB - \tenB^*\|_1 \le u_{\xi}\alpha_\xi \cdot \sqrt{d^*}\twonorm{\htenB - \tenB^*}.\label{eq:Delta-num-bound}
\end{align}
Then we can apply the result of Theorem~\ref{thm:error-bound-Bernoulli} to further bound~\eqref{eq:Delta-num-bound} with high probability.

For the denominator of the RHS of~\eqref{eq:Delta-upper-bound}, we can lower bound it first as $n_{cal}l_\xi$, and for $n_{cal}$, since each tensor entry can become a calibration point independently with probability $\exp[-2h(\tenB_s)](1-q)$, where $0<q<1$ is the train-calibration set split probability, we can then apply the Chernoff bound and obtain:
\begin{equation}\label{eq:ncal-bound}
    \Prob\left(n_{cal} \le (1-c)(1-q)\|\exp[-2h(\tenB^*)]\|_1\right) \le \exp\left[-\frac{c^2(1-q)\|\exp[-2h(\tenB^*)]\|_1}{2}\right],
\end{equation}
for any $0<c<1$. By denoting the event $\{n_{cal} \ge (1-c)(1-q)\|\exp[-2h(\tenB^*)]\|_1\}$ as $\mathcal{E}_0$ and the event in~\eqref{eq:estimator-error-bound} as $\mathcal{E}_1$ and noticing that $\|\exp[-2h(\tenB^*)]\|_1\ge d^*l_\xi$, then we have:
\begin{equation*}
    \Prob\left(\Delta \le \frac{2C_{K,1}}{(1-c)(1-q)}\cdot\frac{u_\xi\alpha_\xi^2}{\gamma_\xi l_\xi^2}\cdot\sqrt{\frac{r^*\bar{d}}{d^*}}\right) \ge 1-\exp[-C_1\bar{d}\log K]-\exp\left[-\frac{c^2(1-q)d^*l_\xi}{2}\right],
\end{equation*}
where the probability is the lower bound of the probability of the event $\mathcal{E}_0\cap\mathcal{E}_1$. With this tail bound on $\Delta$, one can upper bound $\mathrm{E}[\Delta]$ as:
\begin{equation}\label{eq:Delta-expectation-bound}
    \mathrm{E}[\Delta] \le \frac{2C_{K,1}}{(1-c)(1-q)}\cdot\frac{u_\xi\alpha_\xi^2}{\gamma_\xi l_\xi^2}\cdot\sqrt{\frac{r^*\bar{d}}{d^*}} + \exp[-C_1\bar{d}\log K] + \exp\left[-\frac{c^2(1-q)d^*l_\xi}{2}\right],
\end{equation}
and thereby completes the proof.
\end{proof}

\begin{remark}\label{rmk:under-coverage-remark}
Under the scenario where $d_1 \asymp \cdots \asymp d_K \asymp O(d)$ and $r_1 \asymp \cdots r_{K-1} \asymp O(r)$, the coverage shortfall in~\eqref{eq:under-coverage-bound}, i.e. the difference between the lower bound in~\eqref{eq:under-coverage-bound} and $(1-\alpha)$,  can be simplified into:
\begin{equation*}
    \frac{c_{K,\xi}}{(1-c)(1-q)}\cdot\sqrt{\left(\frac{r}{d}\right)^{K-1}} + \exp\left[-c_{K}d\right] + \exp\left[-c^{\prime}_{K,\xi}c^2(1-q)d^K\right],
\end{equation*}
where $c_{K,\xi},c^{\prime}_{K,\xi}$ are positive constants that only relate to $K$ and $\xi$. The first term is of polynomial order with respect to $r/d$ while the other two terms are of exponential order with respect to $d$, therefore the first term is the dominating term and the under-coverage of the conformal intervals scale primarily with $r/d$.
\end{remark}

\section{Technical Lemmas}\label{app:lemma}
All technical lemmas listed in this section are cited from existing works. Therefore, we omit the proof here and refer our readers to the corresponding papers cited.
\begin{lemma}[Lemma 1 of~\citet{wang2020learning}]\label{lemma:inner-product-bound}
For two tensors $\tenA,\tenB\in\mathbb{R}^{\dk}$, their inner product $\inner{\tenA}{\tenB}$ can be bounded as:
\begin{equation*}
    \left|\left\langle\tenA,\tenB\right\rangle\right| \le \|\tenA\|_{\sigma}\|\tenB\|_{*},
\end{equation*}
where $\|\cdot\|_{\sigma},\|\cdot\|_{*}$ are the tensor spectral norm and the tensor nuclear norm, respectively.
\end{lemma}
\begin{lemma}[Lemma 24 of~\citet{cai2022provable}]\label{lemma:rank-bound-tensor-train}
Let $\tenA,\tenB\in\mathbb{R}^{\dk}$ be two low tensor-train rank tensors with $\ranktt{\tenA}=\vecr_1, \ranktt{\tenB}\le\vecr_2$, respectively. Then one has:
\begin{equation*}
    \ranktt{\tenA+\tenB} \le \vecr_1 + \vecr_2.
\end{equation*}
\end{lemma}
\begin{lemma}[Lemma 25 of~\citet{cai2022provable}]\label{lemma:nuclear-frobenius-bound}
Let $\tenA\in\mathbb{R}^{\dk}$ be a low tensor-train rank tensor with $\ranktt{\tenA}=\vecr=(r_1,\ldots,r_{K-1})$ and has a left-orthogonal representation $\tenA = [\tenT_1,\ldots,\tenT_K]$, then:
\begin{equation*}
    \|\tenA\|_* \le \sqrt{r_1\cdots r_{K-1}}\cdot\twonorm{\tenA}.
\end{equation*}
\end{lemma}
\begin{lemma}[Theorem 1 of~\citet{tomioka2014spectral}]\label{lemma:concentration-tensor-spec-norm}
For a random tensor $\tenA\in\mathbb{R}^{\dk}$ with mean-zero and independent sub-Gaussian entries with sub-Gaussian norm $\sigma$, its spectral norm satisfies:
\begin{equation*}
    \|\tenA\|_\sigma \le \sqrt{8\sigma^2\left[\bar{d}\log 5K + \log\frac{2}{\delta}\right]},
\end{equation*}
with probability at least $1-\delta$.
\end{lemma}

\section{Appendix for Simulation}\label{app:simulation}

\subsection{Details of Simulation Setup}\label{app:simulation-details}
We summarize the data-generating model of all essential tensors involved in the simulation experiment in Table~\ref{tab:simulation-details}.
\begin{table}[!htb]
    \centering
    \begin{tabular}{|c|c|c|}
    \hline
    Tensor & Generating Model & Additional Details \\\hline
    $\tenB^*$ & \makecell{$\tenB^* = \tenC\times_1\matU_1\times_2\matU_2\times_3\matU_3$ \\ $\tenC\in\mathbb{R}^{r\times r\times r}, \matU_i\in\mathbb{R}^{d_i\times r_i}$} &  \makecell{$\tenC\overset{i.i.d.}{\sim} 0.5\cdot\mathcal{N}(-1,0.5) + 0.5\cdot\mathcal{N}(1,0.5)$ \\ $\matU_i = \begin{bmatrix}
        1 & \cdots & 1 & 0 & \cdots & 0 & \cdots & 0 & \cdots & 0\\
        0 & \cdots & 0 & 1 & \cdots & 1 & \cdots & 0 & \cdots & 0 \\
        \vdots & \ddots & \vdots & \vdots & \ddots & \vdots & \cdots & \vdots & \ddots & \vdots \\
        0 & \cdots & 0 & 0 & \cdots & 0 & \cdots & 1 & \cdots & 1
    \end{bmatrix}^{\top}$ \\
    and each row of $\matU_i$ has $\left\lceil d_i/r_i\right\rceil$ ones.} \\\hline 
    $\tenW$ & \makecell{$p(\tenW) \propto \exp\left[-\mathcal{H}(\tenW|\tenB^*)\right]$ \\ based on~\eqref{eq:boltzmann-distribution} and~\eqref{eq:hamiltonian}} & \makecell{simulate by block-Gibbs MCMC, \\ where in each proposal we first sample \\ $\mathbb{I}_1=\{(i_1,\ldots,i_K)|\sum_k i_k\text{ is odd}\}$ then $\mathbb{I}_1^{c}$. \\ Each block is a Bernoulli model.} \\\hline
    $\tenX$ & $\tenX = \tenX^* + \tenE$ & $\tenX$ is then masked by $\tenW$. \\\hline
    $\tenX^*$ & \makecell{$\tenX^* = \tenC^*\times_1\matU^*_1\times_2\matU^*_2\times_3\matU^*_3$ \\ $\tenC^*\in\mathbb{R}^{3\times 3\times 3}, \matU^*_i\in\mathbb{R}^{d_i\times 3}$} & $\tenC^*,\matU_1^*,\matU_2^*,\matU_3^* \overset{i.i.d.}{\sim}\mathcal{N}(0,1)$ \\\hline
    $\tenE$ & $[\tenE]_s\overset{\text{independent}}{\sim}\mathcal{N}(0,\sigma_s^2)$ & $\sigma_s = \begin{cases}
        1 & \text{constant noise} \\
        0.5\left[1+\exp(-\tenB^*_s)\right] & \text{adversarial noise}
    \end{cases}$ \\
    \hline
    \end{tabular}
    \caption{Details of the tensors generated in the simulation experiment.}
    \label{tab:simulation-details}
\end{table}

\subsection{Results on the Missing Propensity Estimation Error}\label{app:decomp-simulation}
We examine here the effectiveness of the RGrad algorithm for recovering the tensor parameter $\tenB^*$ from a single observation $\tenW$. We consider $d\in\{40,60,80,100\}$ and $r\in\{3,5,7,9\}$ when simulating $\tenB^*$. For simulating $\tenW$ using the Ising model, we fix $h(x)=x/2$ and consider either $g(x,y)\in\{0, xy/15\}$, where we term the case with $g=0$ as the (independent) Bernoulli model and the case with $g(x,y)=xy/15$ as the (product) Ising model. We split the training and calibration set randomly based on a $70\%-30\%$ ratio.

Under each combination of the choices of $(d,r,g)$, we generate $n=30$ repetitions from a single chain of MCMC and fit RGrad to each repetition with the correctly specified $g(\cdot,\cdot)$ and a working rank $r^{'}\in\{2,3,\ldots,15\}$. In Table~\ref{tab:Model-Selection}, we present the average rank selected by the P-AIC and P-BIC under the Bernoulli and Ising models with various $(d,r)$ combinations.
\begin{table}[!htb]
    \centering
    \begin{tabular}{|c|cccc|cccc|}
    \hline
    \multicolumn{9}{|c|}{Bernoulli Model ($g(x,y)=0$)} \\\hline
    & \multicolumn{4}{c|}{P-AIC} & \multicolumn{4}{c|}{P-BIC} \\\hline
    rank & $d=40$ & $d=60$ & $d=80$ & $d=100$ & $d=40$ & $d=60$ & $d=80$ & $d=100$ \\\cline{2-9}
    $r=3$ & $\textbf{3.0}$ & $\textbf{3.0}$ & $\textbf{3.0}$ & $\textbf{3.0}$ & $2.0$ & $2.0$ & $\textbf{3.0}$ & $\textbf{3.0}$ \\
    $r=5$ & $\textbf{5.0}$ & $\textbf{5.0}$ & $\textbf{5.0}$ & $\textbf{5.0}$ & $2.0$ & $2.1(0.3)$ & $4.0$ & $\textbf{5.0}$ \\
    $r=7$ & $6.2(0.4)$ & $\textbf{7.0}$ & $\textbf{7.0}$ & $\textbf{7.0}$ & $2.0$ & $2.0$ & $2.0$ & $2.3(0.4)$ \\
    $r=9$ & $6.0(0.8)$ & $\mathbf{8.8(0.4)}$ & $\textbf{9.0}$ & $\textbf{9.0}$ & $2.0$ & $2.0$ & $2.0$ & $2.0$ \\
    \hline
    \multicolumn{9}{|c|}{Ising Model ($g(x,y)=xy/15$)} \\\hline
    & \multicolumn{4}{c|}{P-AIC} & \multicolumn{4}{c|}{P-BIC} \\\hline
    rank & $d=40$ & $d=60$ & $d=80$ & $d=100$ & $d=40$ & $d=60$ & $d=80$ & $d=100$ \\\cline{2-9}
    $r=3$ & $\mathbf{3.4(2.0)}$ & $\textbf{3.0}$ & $\textbf{3.0}$ & $\textbf{3.0}$ & $2.0$ & $\textbf{3.0}$ & $\textbf{3.0}$ & $\textbf{3.0}$ \\
    $r=5$ & $\mathbf{7.7(4.1)}$ & $\textbf{5.0}$ & $\textbf{5.0}$ & $\textbf{5.0}$ & $2.0$ & $4.0$ & $\textbf{5.0}$ & $\textbf{5.0}$ \\
    $r=7$ & $13.9(0.2)$ & $\textbf{7.0}$ & $\textbf{7.0}$ & $\textbf{7.0}$ & $2.0$ & $2.1(0.2)$ & $4.0(0.2)$ & $4.7(0.4)$ \\
    $r=9$ & $13.9(0.2)$ & $\textbf{9.0}$ & $\textbf{9.0}$ & $\textbf{9.0}$ & $2.0$ & $2.0$ & $2.0$ & $3.9(0.3)$ \\
    \hline
    \end{tabular}
    \caption{Model selection result of the Bernoulli model and Ising model. Each number is the mean rank selected by P-AIC/P-BIC with $n=30$ repetitions followed by its standard deviations, if non-zero. Boldface is the cases where the true rank is within 1.96 standard deviations of the average rank.}
    \label{tab:Model-Selection}
\end{table}

Based on these numerical results, we find that the consistency of P-AIC and P-BIC depends on $r/d$, or the ``low-rankness" $\tenB^*$. For tensors with high $d$ and low $r$, both P-AIC and P-BIC are consistent, and the inconsistency emerges as $r/d$ becomes larger. Generally speaking, P-AIC is more robust than P-BIC and is consistent across most of the simulation scenarios except for two cases with small tensor sizes. We therefore suggest using P-AIC for rank selection.

\begin{figure}[!htb]
    \centering
    \includegraphics[width=0.98\textwidth]{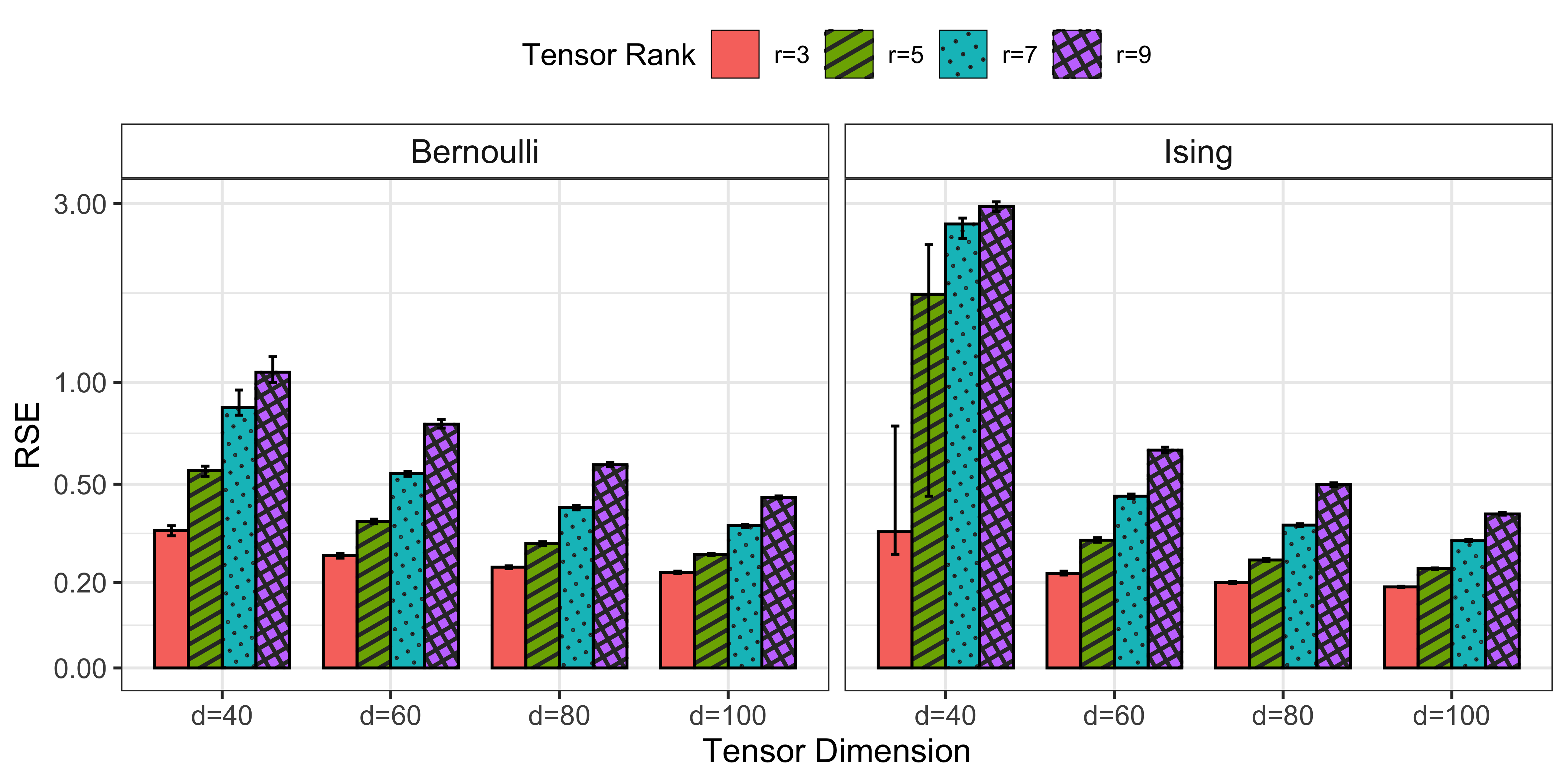}
    \caption{Relative square error of the MPLE $\htenB$ under the Bernoulli (left) and Ising model (right). The results are based on $n=30$ repetitions with the working rank of each sample determined by P-AIC, and each model is fitted by a randomly chosen $70\%$ training set. Error bars show the $2.5\%$ and $97.5\%$ quantiles.}
    \label{fig:estimation-error}
\end{figure}

We then evaluate the fitted $\htenB$ with relative squared error (RSE) defined as: $\twonorm{\htenB-\tenB^*}/\twonorm{\tenB^*}$. The results, as plotted in Figure~\ref{fig:estimation-error}, exhibit a tendency that as $r/d$ becomes larger, so does the RSE, which echoes the results of the model selection. Additionally, the estimation error is lower for the Ising model, as compared to the Bernoulli model, given the same $r$ and $d$. We interpret this result as the Ising model estimator can leverage the additional information from neighbors to infer the missing propensity of each tensor entry. In Figure~\ref{fig:simulation-setup}(d) of the main paper, we plot the estimator for $\tenB^*$ shown in~\ref{fig:simulation-setup}(a) by RGrad based on a randomly chosen $70\%$ training set, and it is clear that $\htenB$ reconstructs $\tenB^*$ very well.

\subsection{Results on Conformal Prediction Validation}\label{app:simulation-results}
\begin{figure}[!htb]
    \centering
    \includegraphics[width=0.98\textwidth]{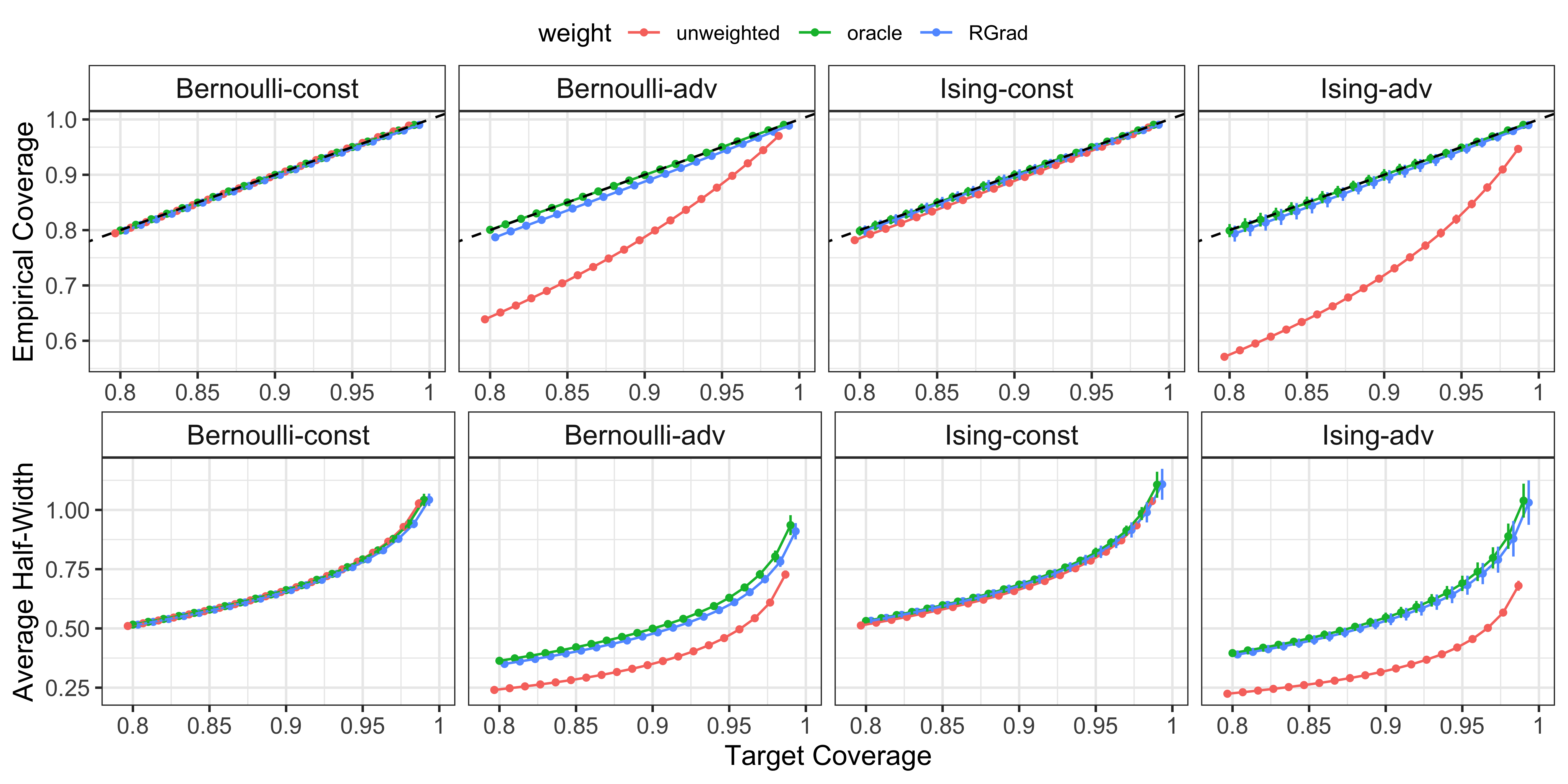}
    \caption{Empirical coverage and average confidence interval half-width of the three conformal prediction methods across the Bernoulli and Ising model with constant (const) or adversarial (adv) noise. Results are based on $n=30$ repetitions, and error bars are $\pm 1.96$ standard deviations.}
    \label{fig:sim-coverage-additional-result}
\end{figure}

As a companion result of Figure~\ref{fig:sim-conformal-coverage}, we plot the empirical coverage and half of the average confidence interval width of three conformal prediction methods under different simulation scenarios in Figure~\ref{fig:sim-coverage-additional-result}. The mis-coverage of the unweighted conformal prediction comes from under-coverage and is associated with shorter confidence intervals. The reason why unweighted conformal prediction has under-coverage is that under the adversarial noise setting, entries with higher missing propensity also have higher uncertainty, and using a uniform weight underestimates the uncertainty of a missing entry. As one can tell from Figure~\ref{fig:sim-coverage-additional-result}, our CTC algorithm matches the oracle case quite well and provides well-calibrated confidence intervals.

Apart from these results, we also compared other binary tensor decomposition methods for estimating the missing propensity and conducting the weighted conformal prediction with our method. We mainly consider two competing methods other than the unweighted and oracle conformal prediction: 1) \textbf{GCP}: binary tensor decomposition with generalized CP-decomposition~\citep{wang2020learning,hong2020generalized}; 2) \textbf{Tucker}: binary tensor decomposition with generalized Tucker-decomposition~\citep{lee2020tensor,cai2022generalized}. Different from our approach, these two methods assume independence among all the binary entries and thus they are misspecified under the Ising model. We conduct \textbf{GCP} with gradient descent following~\citet{hong2020generalized} and \textbf{Tucker} with Riemannian gradient descent following~\citep{cai2022generalized} and select the corresponding ranks using the BIC criterion, as suggested by the literature. We consider $r=3, d\in\{40,60\}$ and list the average mis-coverage \% under the constant and adversarial noise regimes as well as the RSE of the estimated $\htenB$ in Table~\ref{tab:conformal_compare_method}.

Our finding from Table~\ref{tab:conformal_compare_method} is that our method consistently provides well-calibrated confidence intervals close to the oracle case and performs, on average, better than the GCP and Tucker method. Our mis-coverage \% is statistically significantly better (p-value $<0.005$) than the Tucker method under the adversarial noise regimes across different tensor dimensions and missingness generating models. The GCP method, surprisingly, provides confidence intervals close to our method but has significantly larger RSE for the estimator $\htenB$. We found that CP-decomposition tends to underestimate the weights of the calibration data; therefore, it has more testing data points with infinitely wide confidence intervals, making it less favorable.
\begin{table}[!htb]
    \centering
    \begin{tabular}{|c|c|c|c|c|}
    \hline
    ($d$, Model) & Method & const. mis-coverage \% & adv. mis-coverage \% & RSE \\\hline
    \multirow{5}{*}{(40, Bern)}  & unweighted & $0.463(0.244)$ & $11.1(0.389)$ & / \\
    & oracle & $0.381(0.205)$ & $0.409(0.232)$ & / \\
    & GCP  & $0.373(0.183)$ & $1.66(0.965)$ & $0.522(0.069)$ \\
    & Tucker & $0.380(0.219)$ & $0.841(0.431)$ & $0.295(0.008)$ \\
    & RGrad & $0.377(0.235)$ &  $0.773(0.404)$ & $0.345(0.010)$ \\\hline
    \multirow{5}{*}{(60, Bern)}  & unweighted & $0.401(0.165)$ & $11.0(0.241)$ & / \\
    & oracle & $0.202(0.082)$ & $0.207(0.105)$ & / \\
    & GCP  & $0.203(0.092)$ & $0.380(0.298)$ & $0.281(0.036)$ \\
    & Tucker & $0.199(0.079)$ & $0.842(0.231)$ & $0.244(0.005)$ \\
    & RGrad & $0.200(0.078)$ &  $0.821(0.226)$ & $0.271(0.004)$ \\\hline
    \multirow{5}{*}{(40, Ising)}  & unweighted & $1.19(0.298)$ & $17.3(0.528)$ & / \\
    & oracle & $0.568(0.278)$ & $0.666(0.331)$ & / \\
    & GCP  & $0.870(0.597)$ & $1.24(0.840)$ & $1.81(0.621)$ \\
    & Tucker & $0.504(0.241)$ & $1.80(0.653)$ & $0.444(0.010)$ \\
    & RGrad & $0.713(0.377)$ &  $1.13(1.21)$ & $\mathbf{0.341(0.304)}$ \\\hline
    \multirow{5}{*}{(60, Ising)}  & unweighted & $1.35(0.243)$ & $17.2(0.310)$ & / \\
    & oracle & $0.302(0.136)$ & $0.370(0.242)$ & / \\
    & GCP  & $0.349(0.181)$ & $0.638(0.506)$ & $1.59(1.16)$ \\
    & Tucker & $0.329(0.216)$ & $2.03(0.368)$ & $0.404(0.007)$ \\
    & RGrad & $0.356(0.154)$ &  $0.580(0.339)$ & $\mathbf{0.224(0.003)}$ \\\hline
    \end{tabular}
    \caption{Method comparisons of different conformal prediction methods with $r=3$. The results include the average mis-coverage \% defined in~\eqref{eq:coverage-l1-loss} under the constant (const.) and adversarial (adv.) noise regimes as well as the relatively squared error (RSE) of the estimator $\htenB$.}
    \label{tab:conformal_compare_method}
\end{table}

\subsection{Results with Misspecified Tensor Completion Model}\label{app:misspec-tc-sim}
\begin{figure}[!htb]
    \centering
    \includegraphics[width=\linewidth]{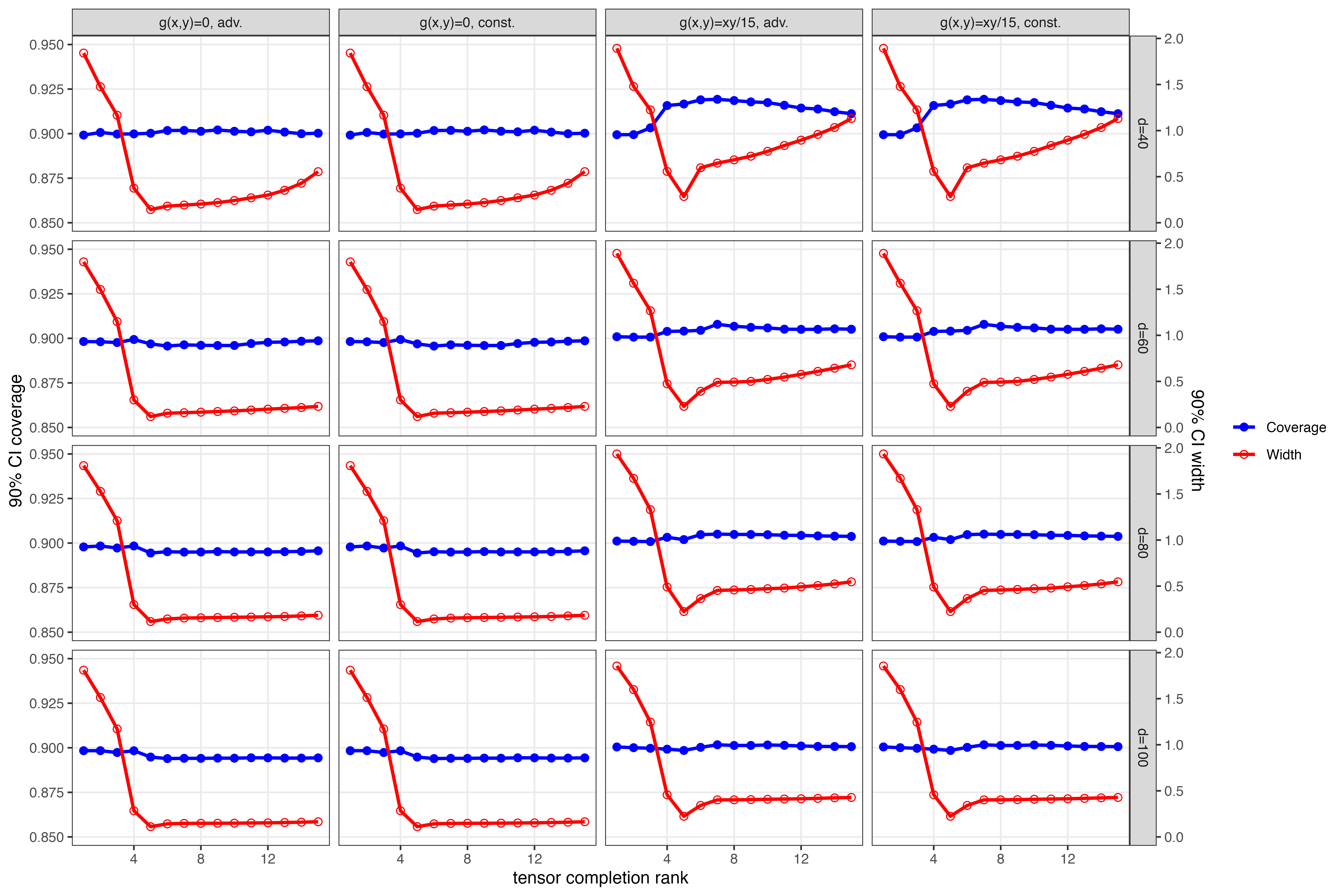}
    \caption{CTC $90\%$ confidence interval coverage and width against different tensor completion model ranks, under different tensor dimensions $d$, additive noise distribution (adv. = adversarial, const. = constant) and missingness correlations (as specified by $g(\cdot, \cdot)$). Results based on 30 repetitions, and $\tenB$ has a fixed rank of $r=3$ and $\htenB$ estimated by Algorithm~\ref{alg:RGrad}.}
    \label{fig:tc_misspecified_sim}
\end{figure}

In this section, we demonstrate the property of our CTC algorithm under a misspecified tensor completion model. We fixate on the scenario where $\tenB^*$ has rank $r=3$, and $g(x,y)\in\{0, xy/15\}$. We make a claim in Proposition~\ref{thm:conformal-interval-weight} that our conformal inference has valid coverage under any \textit{arbitrary} choice of the tensor completion algorithm that generated $\htenX$. To exhibit the impact of model misspecification on the conformal inference, we generate the data tensor following $\tenX=\tenX^* + \tenE$, with $\tenX^*$ having a Tucker rank at $(5,5,5)$ and $\tenE$ following either constant or adversarial noise (see Section~\ref{subsec:conformal-simulation} for detailed definition). Then we run a tensor completion algorithm with a working rank of $(k, k, k), k\in[1, 15]$, and we compute the average coverage and confidence interval width for the $90\%$ CI, where the CI is generated with our CTC algorithm with $\htenB$ estimated using RGrad. We report the results under tensor dimension $d\in\{40, 60, 80, 100\}$ in Figure~\ref{fig:tc_misspecified_sim}.

In line with our expectations, we observe that the coverage probability remains well-calibrated at $90\%$ across all simulation scenarios, regardless of the tensor completion rank. An exception is when missingness is correlated $g(x,y)=xy/15$ and when the tensor dimension is low at $d=40$, which is not surprising given that the estimation error for $\htenB$ is higher. Also, we notice that the CI width is optimal under the correct rank ($k=5$), and becomes slightly worse when one overspecifies the model and significantly worse when one underspecifies the model. The conclusion is that our CTC algorithm has well-calibrated coverage probability and will have sub-optimal CI width when one misspecifies the tensor completion model. Given that the tensor completion model can properly select the rank using model selection criteria like AIC/BIC, we consider this a minor issue for our CTC algorithm.

\subsection{Results on Other Non-conformity Scores}\label{app:ncs-sim}
In this section, we expand the experiments conducted in Section~\ref{subsec:conformal-simulation} by considering other choices of the non-conformity scores. We primarily consider two additional types of scores. Firstly, the two-sided non-conformity score, which is basically $\mathcal{S}(\tenX_s, \htenX_s) = \tenX_s - \htenX_s$. For any $s^*\in\setS_{miss}$, similar to~\eqref{eq:qhat-weighted-definition}, we define $\qhat_{s^*, l}$ and $\qhat_{s^*, r}$ as:
\begin{equation*}
    \qhat_{s^*, l} = \mathcal{Q}_{(1-\alpha)/2} \left(\sum_{i=1}^{n} \omega_{i}(s^*) \cdot \delta_{\mathcal{S}(\tenX_{s_i},\htenX_{s_i})} + \omega_{n+1}(s^*)\cdot\delta_{+\infty}\right),\quad \text{where }\omega_k(s^*) = \frac{p_k(s^*)}{\sum_{i=1}^{n+1} p_i(s^*)},
\end{equation*}
\begin{equation*}
    \qhat_{s^*, r} = \mathcal{Q}_{(1+\alpha)/2} \left(\sum_{i=1}^{n} \omega_{i}(s^*) \cdot \delta_{\mathcal{S}(\tenX_{s_i},\htenX_{s_i})} + \omega_{n+1}(s^*)\cdot\delta_{+\infty}\right),\quad \text{where }\omega_k(s^*) = \frac{p_k(s^*)}{\sum_{i=1}^{n+1} p_i(s^*)},
\end{equation*}
and construct the $(1-\alpha)$-level conformal interval as $C_{1-\alpha,s^*}(\htenX) = \{x\in\mathbb{R}|\qhat_{s^*,l} \le \mathcal{S}(x,\htenX_{s^*}) \le \qhat_{s^*,r}\}$.

The second type of score is the normalized non-conformity score, which is defined as $\mathcal{S}(\tenX_s, \htenX_s) = |\tenX_s - \htenX_s| / \uhat_s$, where $\uhat_s$ is a prior estimate of the uncertainty at $s$. The $\uhat_s$ can be based on the uncertainty quantification of tensor completion estimators. Typically for matrix/tensor completion work with uncertainty quantification~\citep{chen2019inference,farias2022uncertainty,gui2023conformalized,ma2024statistical}, all entries are assumed to be missing independently with the same probability, which makes it harder to directly utilize the asymptotic normality of the estimator for quantifying $\widehat{u}_s$ under our context. In this section, we use the entrywise confidence interval for tensor completion with low Tucker rank, which is also the completion algorithm we use, from~\citet{ma2024statistical} (see Corollary 1). For each $\widehat{u}_s$, it is proportional to $\twonorm{\proj{\setT}{\mathbfcal{I}_s}}$, where $\setT$ is the tangent space of $\htenX$ and $\mathbfcal{I}_s$ is a binary tensor with all but one entry being 0, and entry $s$ being 1. This is essentially a misspecified model since our data are not missing uniformly at random. The $(1-\alpha)$-level conformal interval is constructed via $C_{1-\alpha,s^*}(\htenX) = \{x\in\mathbb{R}| \left|\tenX_{s^*} - \htenX_{s^*}\right| \le \qhat_{s^*} \cdot \uhat_{s^*}\}$, where $\qhat_{s^*}$ is defined similarly as~\eqref{eq:qhat-weighted-definition}, with the non-conformity score being the normalized score. 

\begin{figure}[h]
    \centering
    \includegraphics[width=0.98\linewidth]{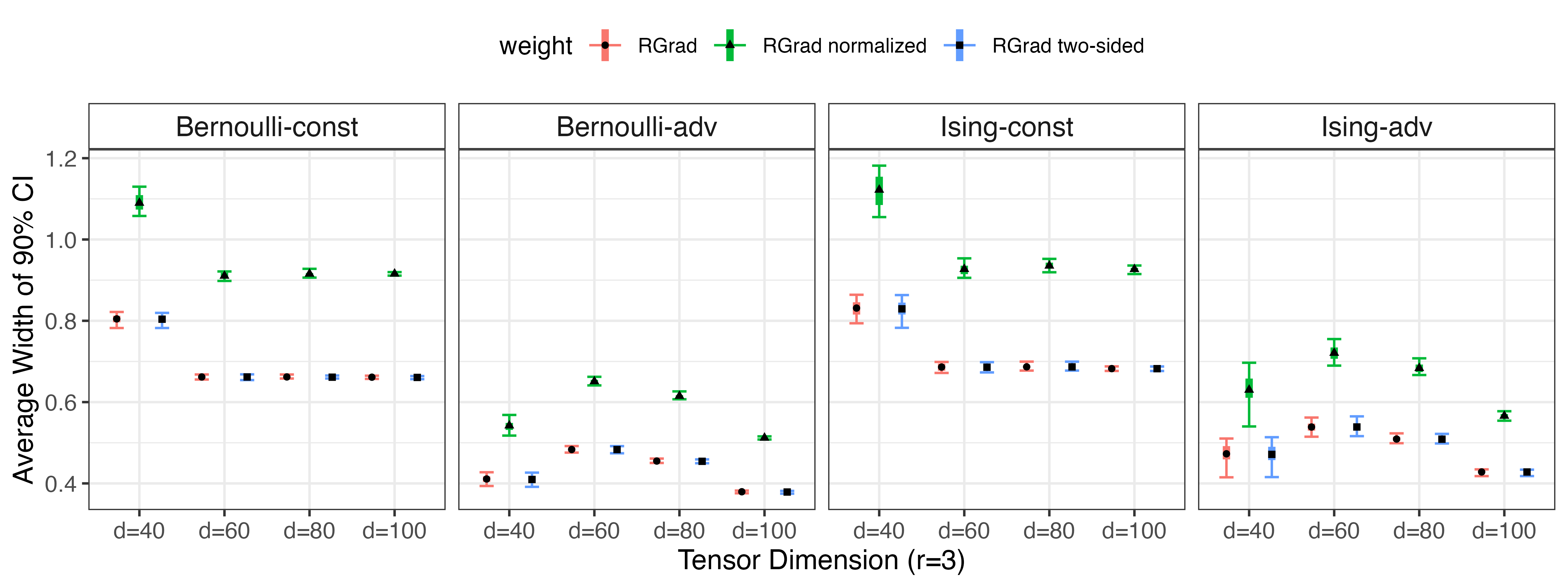}
    \caption{The average width of the $90\%$ conformal interval of conformal prediction methods under three different non-conformity scores, with $d\in\{40,60,80,100\}$, $r=3$ under the Bernoulli and Ising model. Two uncertainty regimes: constant noise (const) and adversarial noise (adv) are considered. Results are based on $30$ repetitions, error bars show the $2.5\%,97.5\%$ quantiles, and the thicker lines show the range of $25\%$ to $75\%$ quantiles.}
    \label{fig:simulation-C90-width}
\end{figure}

We re-run the experiment in Section~\ref{subsec:conformal-simulation} and summarize the average mis-coverage\%, as defined in~\eqref{eq:coverage-l1-loss}, in Table~\ref{tab:ncs-compare}. From Table~\ref{tab:ncs-compare}, we saw generally no significant gain in using either score. Given that the prior uncertainty estimates $\uhat_s$ are under a misspecified model, we even see some worse coverage when using the normalized score. In Figure~\ref{fig:simulation-C90-width}, we also plotted the average width of the $90\%$ confidence interval under un-normalized, normalized, and two-sided non-conformity scores. The impact of misspecifying the prior uncertainty estimate is making the confidence interval wider. Overall, our results suggest that the miscoverage is mild under any of these non-conformity score definitions, which also showcases the robustness of our approach.
\begin{table}[!h]
\centering
\begin{tabular}{|ccccccc|}
\toprule
Model & $d$ & Unweighted & Oracle & RGrad & \parbox{2cm}{\centering RGrad \\ Normalized} & \parbox{2cm}{\centering RGrad \\ Two-sided}\\
\midrule
\multicolumn{7}{|c|}{Constant Noise} \\\hline
\multirow{4}{*}{Bernoulli} & d=40 & 0.46 (0.24) & 0.38 (0.21) & 0.38 (0.23) & 0.44 (0.25) & 0.38 (0.23)\\
 & d=60 & 0.4 (0.17) & 0.2 (0.08) & 0.2 (0.08) & 0.21 (0.11) & 0.2 (0.08)\\
 & d=80 & 0.44 (0.1) & 0.12 (0.06) & 0.12 (0.06) & 0.12 (0.06) & 0.12 (0.06)\\
 & d=100 & 0.37 (0.08) & 0.09 (0.06) & 0.09 (0.06) & 0.09 (0.04) & 0.09 (0.06)\\\hline
\multirow{4}{*}{Ising} & d=40 & 1.19 (0.3) & 0.57 (0.28) & 0.71 (0.38) & 0.82 (0.41) & 0.71 (0.38)\\
 & d=60 & 1.35 (0.24) & 0.3 (0.14) & 0.36 (0.15) & 0.36 (0.17) & 0.36 (0.15)\\
 & d=80 & 1.4 (0.11) & 0.22 (0.12) & 0.21 (0.15) & 0.23 (0.14) & 0.21 (0.15)\\
 & d=100 & 1.26 (0.09) & 0.12 (0.07) & 0.12 (0.07) & 0.16 (0.11) & 0.12 (0.07)\\\hline
\multicolumn{7}{|c|}{Adversarial Noise} \\\hline
\multirow{4}{*}{Bernoulli} & d=40 & 11.12 (0.39) & 0.41 (0.23) & 0.77 (0.4) & 0.73 (0.46) & 0.79 (0.41)\\
 & d=60 & 11.05 (0.24) & 0.21 (0.1) & 0.82 (0.23) & 0.72 (0.24) & 0.83 (0.23)\\
 & d=80 & 11.24 (0.14) & 0.14 (0.07) & 0.89 (0.15) & 0.82 (0.16) & 0.89 (0.15)\\
& d=100 & 10.81 (0.12) & 0.1 (0.07) & 0.81 (0.12) & 0.74 (0.1) & 0.81 (0.12)\\\hline
\multirow{4}{*}{Ising} & d=40 & 17.31 (0.53) & 0.67 (0.33) & 1.13 (1.21) & 1.17 (0.83) & 1.06 (1.21)\\
 & d=60 & 17.23 (0.31) & 0.37 (0.24) & 0.58 (0.34) & 0.51 (0.28) & 0.59 (0.35)\\
 & d=80 & 17.61 (0.2) & 0.25 (0.14) & 0.5 (0.26) & 0.43 (0.27) & 0.51 (0.25)\\
 & d=100 & 17.08 (0.12) & 0.17 (0.12) & 0.53 (0.2) & 0.49 (0.19) & 0.53 (0.19)\\
\bottomrule
\end{tabular}
\caption{Average mis-coverage\% (and standard deviation, also in percentage) for unweighted, oracle, RGrad, RGrad with normalized non-conformity score, and RGrad with two-sided non-conformity score. We set $r=3$ in all cases, and the results are based on 30 iterations.}
\label{tab:ncs-compare}
\end{table}

\subsection{Algorithm Stepsize Selection, Runtime and Convergence}\label{app:time_n_conv}
\begin{figure}[h]
    \centering
    \includegraphics[width=0.98\linewidth]{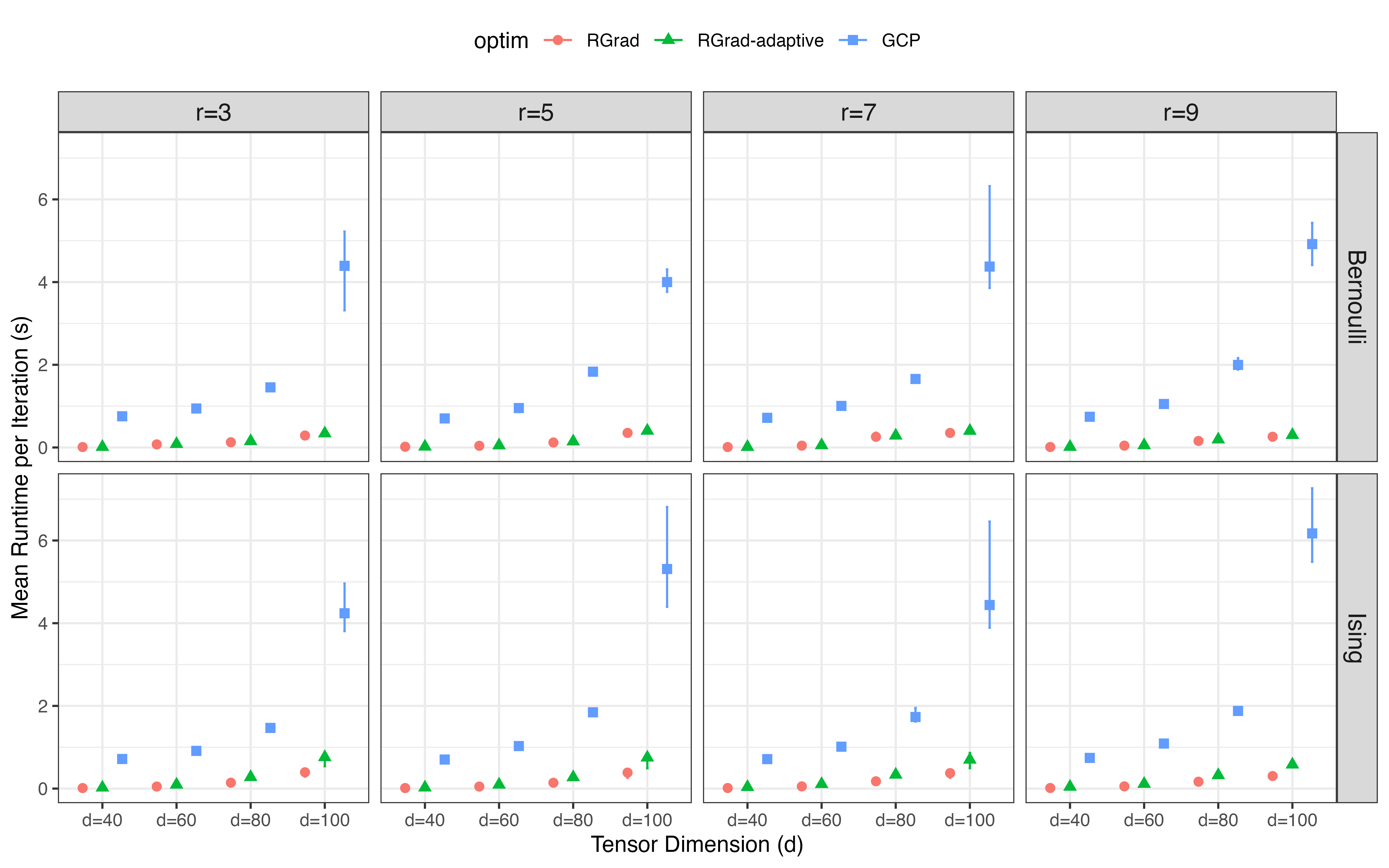}
    \caption{Runtime in seconds per iteration comparison. All three algorithms are run on a 2x 3.0 GHz Intel Xeon Gold 6154 CPU. Errorbar show the $2.5\%$ and $97.5\%$ quantile across 30 iterations.}
    \label{fig:timing_experiment}
\end{figure}

In this subsection, we further explore the convergence, running time, and step size choice of the Riemannian Gradient descent algorithm. In Algorithm~\ref{alg:RGrad}, we set a fixed step size $\eta=0.1$, which can be quite restrictive. Here, we explore an alternative step size scheme based on linearized line search. Specifically, at iteration $(l+1)$, we start with a relatively large stepsize $\eta^\prime$, and check if the following Armijo condition holds for a prespecified hyper-parameter $\alpha$:
\begin{equation}
\ell\left(\tenW_{tr}|\tenB_l\right) - \ell\left(\tenW_{tr}|\TTSVD{\tenB_l-\eta^\prime\proj{\setT_l}{\tenG_l}}\right) \ge \alpha \cdot \eta^\prime\twonorm{\proj{\setT_l}{\tenG_l}}^2.  
\end{equation}
If it does not hold, we set $\eta^\prime \gets \eta^\prime/2$ and continue checking. Basically, we are selecting a step size that reaches sufficient descent of the target negative pseudo-likelihood $\ell(\cdot)$. In Figure~\ref{fig:timing_experiment}, we compare the runtime per iteration of our Algorithm~\ref{alg:RGrad} with fixed $\eta=0.1$ (RGrad), Algorithm~\ref{alg:RGrad} with adaptive step size (RGrad-adaptive), and the binary tensor decomposition with generalized CP-decomposition~\citep{wang2020learning,hong2020generalized} (GCP), where the GCP is essentially based on a gradient descent algorithm. For GCP, we use the MATLAB Tensor Toolbox for implementation and set the CP rank such that the number of parameters is roughly the same as our tensor-train decomposition parameter.

We see that the RGrad-based algorithm is much faster than the GCP per iteration. It is also our empirical observation that RGrad-adaptive requires a much smaller number of iterations on average than the other two algorithms, indicating the potential of further speeding up our algorithm.

\begin{figure}[h]
    \centering
    \includegraphics[width=0.98\linewidth]{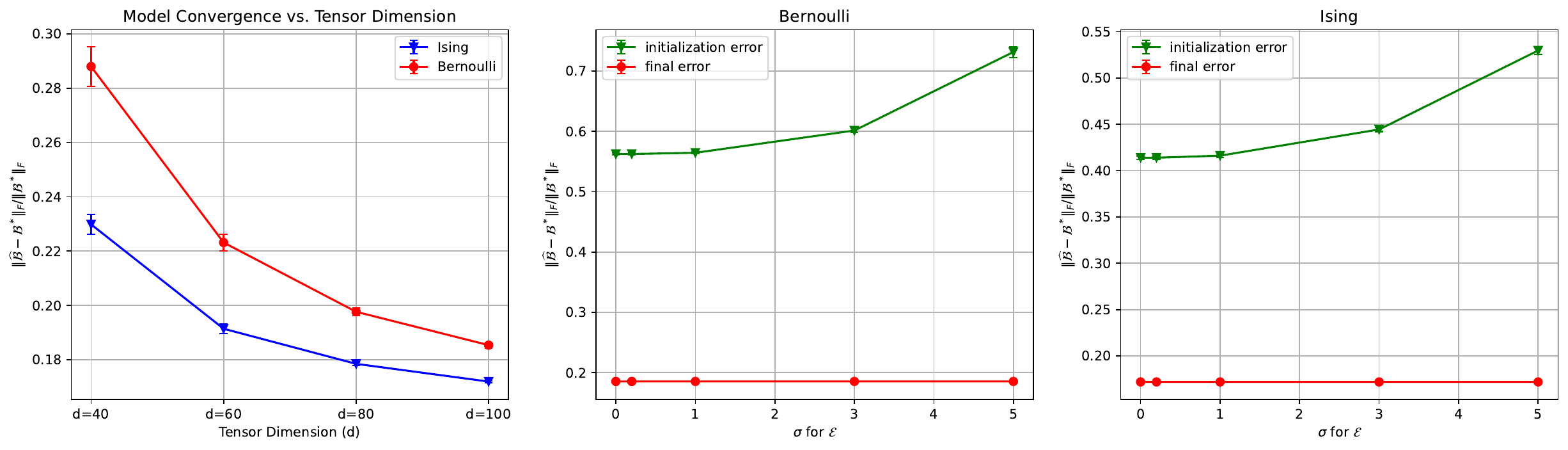}
    \caption{Left: Final estimator RSE vs. Tensor Dimension for RGrad. Middle \& Right: Initialization and Final RSE of the estimator across various levels of random perturbation of the binary tensor during spectral initialization. All results based on 30 iterations and confidence bands are $\pm 1.96$ std.}
    \label{fig:conv_robust}
\end{figure}

We further check the convergence of our RGrad algorithm and the robustness of our initialization method. For convergence, we fit our RGrad under $d\in\{40,60,80,100\}$ with $r=3$, and measure the relative squared error (RSE) of $\htenB$ across both Bernoulli and Ising models. The RSE is simply $\twonorm{\htenB - \tenB^*} / \twonorm{\tenB^*}$. We show the result in the left panel of Figure~\ref{fig:conv_robust}. We generally see that the higher the tensor dimension, the lower the error is. There is an error bound for the RSE, and generally, one requires a higher dimension relative to rank to converge to the true tensor $\tenB^*$. In the middle and right panel of Figure~\ref{fig:conv_robust}, we plot the initialization RSE and final RSE of the RGrad algorithm under various standard deviations of the error $\tenE$ that we apply to the binary tensor $\tenW$ for the spectral initialization. We found that regardless of the initialization error, the final error is the same, and thus we could simply do a low-rank TT-SVD over $\tenW$ as initialization.
\end{document}